\documentclass[traditabstract]{aa} 
\usepackage{amsmath}
\usepackage{amssymb}
\usepackage{enumerate}

\usepackage{graphicx}
\usepackage{txfonts}
\usepackage{natbib}

\usepackage{array,tabularx}
\usepackage{booktabs}

\usepackage{colortbl}

\bibpunct{(}{)}{;}{a}{}{,} 

\begin{document}

   \title{Mass distributions of star clusters for different\\ star formation histories in a galaxy cluster environment}

   \author{C. Schulz \inst{1} \and J. Pflamm-Altenburg \inst{2} \and P. Kroupa \inst{2}}

   \institute{  \inst{1} European Southern Observatory (ESO), Karl-Schwarzschild-Stra{\ss}e 2, D-85748 Garching, Germany \newline
                \inst{2} Helmholtz-Institut f\"ur Strahlen- und Kernphysik (HISKP), Universit\"at Bonn, Nussallee 14 - 16, D-53115 Bonn, Germany \newline
                \email{cschulz@eso.org, [jpflamm;pavel]@astro.uni-bonn.de}
             }

\abstract{
Clusters of galaxies usually contain rich populations of globular clusters (GCs). We investigate how different star formation histories (SFHs) shape the final mass distribution of star clusters.

We assumed that every star cluster population forms during a formation epoch of length $\delta t$ at a constant star-formation rate (SFR). The mass distribution of such a population is described by the embedded cluster mass function (ECMF), which is a pure power law extending to an upper limit $M_{\mathrm{max}}$. Since the SFR determines $M_{\mathrm{max}}$, the ECMF implicitly depends on the SFR.

Starting with different SFHs, the time-evolution of the SFR, each SFH is divided into formation epochs of length $\delta t$ at different SFRs. The requested mass function arises from the superposition of the star clusters of all formation epochs. An improved optimal sampling technique is introduced that allows generating number and mass distributions, both of which accurately agree with the ECMF. Moreover, for each SFH the distribution function of all involved SFRs, $F(\mathrm{SFR})$, is computed. For monotonically decreasing SFHs, we found that $F(\mathrm{SFR})$ always follows a power law.

With $F(\mathrm{SFR})$, we developed the theory of the integrated galactic embedded cluster mass function (IGECMF). The latter describes the distribution function of birth stellar masses of star clusters that accumulated over a formation episode much longer than $\delta t$. The IGECMF indeed reproduces the mass distribution of star clusters created according to the superposition principle. Interestingly, all considered SFHs lead to a turn-down with increasing star cluster mass in their respective IGECMFs in a similar way as is observed for GC systems in different galaxy clusters, which offers the possibility of determining the conditions under which a GC system was assembled.

Although assuming a pure power-law ECMF, a Schechter-like IGECMF emerges from the superposition principle. In the past decade, a turn-down at the high-mass end has been observed in the cluster initial mass function. This turn-down can be explained naturally if the observed star cluster ensembles are superpositions of several individual star cluster populations that formed at different times at different SFRs.
}

\date{\today}
\keywords{galaxies: clusters: general -- galaxies: star clusters: general -- methods: analytical}

\maketitle

\section{Introduction} \label{intro}

Galaxy clusters form through the coalescence of galaxy groups and the infall of individual galaxies. In the course of time, the number of galaxies in a galaxy cluster steadily grows, thereby increasing the density of galaxies and enlarging the probability for galaxy-galaxy encounters during which star formation takes place. Thus, galaxy clusters and their major galaxies were assembled from many such encounters in which new stars and star clusters (SCs) were formed, as well as through the accretion of gas, stars, and SCs.

During such collisions, star formation is expected to occur at a higher rate than observed in today's Universe since the early galaxies were notably gas-rich. In these events, SCs of a wide mass range were formed, and due to the high star-formation rates (SFRs), a substantial number of high-mass SCs were able to form \citep[e.g.,][]{larsen_richtler00}. Even if massive stars have short lifetimes and low-mass SCs dissolve fast in a tidal field, at least the high-mass SCs have a considerable chance to survive over a Hubble time, enabling us to observe them today as globular clusters (GCs).

Our idea is to use the surrounding SCs -- in this case, the ancient GCs that probably formed during the above-mentioned interactions -- to derive the star formation activities at that time and to determine under which conditions the major galaxies, and in the end, the host galaxy cluster itself, were assembled. The overall mass distribution of these GCs may show features that bear a memory of such events because in galaxy-galaxy interactions SC populations are formed during bursty phases, while infalling galaxies contribute SC populations that largely formed under quiescent conditions. Other collaborations also use SCs to constrain properties of the host galaxy \citep[e.g.,][]{cote98, maschberger07, beasley08, norris_kannappan11, georgiev12}. To this end, it is investigated how the star formation activities influence the mass distribution of SCs to provide the theoretical groundwork for analyzing SC systems in the future.

What shapes the mass distribution of SCs? Since the major galaxies in a galaxy cluster have probably undergone several interaction processes with intense star formation episodes in the past, an SC sample observed around such a galaxy is most likely a superposition of different SC populations formed at different times. From SC formation it is known that the mass distribution function of a newly born SC population that formed coevally can be described by the so-called embedded cluster mass function (ECMF) \citep[e.g.,][]{lada_lada03, kroupa_weidner03, weidner_kroupa05}. We here assumed that SC formation occurred information epochs of length $\delta t$ at a constant SFR during which the ECMF is fully populated.

In the following, we derive the overall mass distribution of SCs for SC populations that formed during a Hubble time of SC formation, for instance, around an individual galaxy. Analogously to the ECMF, this mass distribution function is called the integrated galactic embedded cluster mass function (IGECMF). It can be obtained by summing the ECMFs of all star-formation epochs that occurred during the different galaxy-galaxy encounters. However, the involved ECMFs are not identical: According to the SFR-$M_{\mathrm{ecl,max}}$ relation \citep{weidner04}, the higher the SFR, the higher the mass of the most massive SC, $M_{\mathrm{ecl,max}}$, and vice versa. Hence, during an SC formation epoch with a high SFR, more and heavier SCs form than within a period with a low SFR. Therefore it is necessary to scale the involved ECMFs by the respective SFR before summing them. Thus, one has to know how the SFRs of the different formation epochs are distributed. This distribution function of all involved SFRs is called $F(\mathrm{SFR})$. Mathematically, to obtain the IGECMF, one has to integrate the ECMF scaled by $F(\mathrm{SFR})$ over all possible SFRs. Physically, one has to sum all SC populations resulting from the different formation epochs over time. 

The distribution function of SFRs, $F(\mathrm{SFR})$, is here quantified from a total of six star formation histories (SFHs): i) two exponential SFHs \citep{gavazzi02} and ii) four power-law SFHs. The latter have not been investigated until now, but by analyzing them, a fundamental relation between an SFH and its $F(\mathrm{SFR})$ is discovered. From the obtained $F(\mathrm{SFR}),
$ we determine the desired IGECMF.

As mentioned previously, the IGECMF describes how the masses of newly born SCs are distributed and must not be confused with the mass distribution function of SCs observed today, the so-called present-day mass function. The reason is that each distribution of SCs will suffer from changes driven by stellar and internal dynamical evolution as well as dissolution and SC disruption due to relaxation, dynamical friction, or tidal effects. Starting with an IGECMF, the present-day mass function can be derived by taking into account all these phenomenons. Since the destructive processes are studied in detail elsewhere (e.g., \citealt{baumgardt_makino03, lamers05, lamers10, lamers13, alexander12, alexander14, gieles11, gieles14, brockamp14}; see also \citealt{portegies_zwart10} and references therein), the subsequent investigations are restricted on the determination of the IGECMFs, which reveal how the birth stellar masses of SCs will theoretically be distributed as a function of different SFHs. To this end, a new optimal sampling technique is developed that allows sampling a population of SCs that ideally resembles the underlying distribution function in terms of the number of objects and their mass.

\subsection{Summarized approach} 
After introducing the necessary background in Sects.~\ref{theory} and \ref{sect_dt}, we develop in Sects.~\ref{sect_SFH} -- \ref{sect_sc_distr} how the theoretical overall mass distribution of SCs is shaped after a star formation period of a particular length. Here we list the main procedure steps and where they are described in detail:

\begin{enumerate}
\item Different SFHs are assumed that reveal how the SFR evolved over a Hubble time (Sect.~\ref{sect_SFH}). Based on the assumption that SC formation occurs in epochs of duration $\delta t$ (Sect.~\ref{subsect_ecmf}, determination in Sect.~\ref{sect_dt}), the SFH can be divided into individual formation epochs each at a particular SFR (Fig.~\ref{fig_sketch}).
\item For each formation epoch the mass of the most massive SC, $M_{\mathrm{ecl,max}}$, can be deduced according to the SFR-$M_{\mathrm{ecl,max}}$ relation (Eq.~\ref{sfr_M_eclmax}, Fig.~\ref{fig_sfr_Mmax5}) that
fully determines the ECMF of that formation epoch (Sect.~\ref{subsect_ecmf}). Thus, the only independent parameter is the SFR that determines all other quantities.
\item The overall mass distribution of SCs after a certain period of SC formation -- here a Hubble time -- emerges from the accumulation of all SC ever formed. This approach is called the superposition principle\footnote{Similarly, \citet{kroupa02} accumulated the velocities of all stars of all embedded SCs in a galaxy to gain their vertical velocity field, while \citet{kroupa_weidner03} obtained the integrated galactic IMF from the masses of these stars.}. Two different methods are used to obtain the final SC mass distribution of birth stellar masses (Sect.~\ref{sect_sc_distr}):
\begin{enumerate}
\item First, the superposition of the ECMFs of all formation epochs modulated by the distribution function of SFRs, $F(\mathrm{SFR})$ (definition in Sect.~\ref{subsect_igecmf}, determination in Sect.~\ref{sect_F}, results in Fig.~\ref{fig_F_20+23}), leads to the IGECMF (definition in Sect.~\ref{subsect_igecmf}, determination in Sect.~\ref{sect_igecmf}, results in Fig.~\ref{fig_igecmf_20+23}). Discretizing the IGECMF gives the overall mass distribution of SCs (continuous lines in Fig.~\ref{fig_nup_20+23}).
\item First, the ECMF of each formation epoch is discretized into one individual SC population using an accurate sampling technique (Sect.~\ref{subsect_sampling}). Then the superposition of all these SC populations leads to the final SC mass distribution (dashed lines in Fig.~\ref{fig_nup_20+23}).
\end{enumerate}
We show that the results from (a) and (b) agree.
\end{enumerate}

To treat this accurately, some mathematical derivations are required (especially Sects.~\ref{subsect_sampling} -- \ref{subsect_igecmf}). To facilitate the understanding of this approach, we recommend reading Sects.~\ref{subsect_exempl} and \ref{subsect_sfh}, in which we explain how the overall star cluster mass function emerges based on different SFHs. This is additionally depicted in Fig.~\ref{fig_intro_plot}, which visualizes how the superposition principle works.

The key finding of this work is that the overall mass function of SCs is expected to exhibit a turn-down at the high-mass end if the SFR changed significantly during the considered SC formation timescale. Since the mass of an SC changes during lifetime, an observed SC mass distribution cannot be directly compared to a theoretical SC mass distribution. Thus, we recommend reading Sect.~\ref{sect_obs}, which summarizes to which extent the predicted shape of the SC mass function can be observed and which corrections are necessary to enable this comparison in order to obtain reasonable results. Finally, we conclude our investigations in Sect.~\ref{sect_concl}.

\section{Underlying theory} \label{theory}

In this section, we first introduce a significantly improved optimal sampling technique in Sect.~\ref{subsect_sampling}. Then we define the ECMF from which the SFR-$M_{\mathrm{ecl,max}}$ relation is derived in Sect.~\ref{subsect_ecmf}. The approach used here is similar to that described by \citet{weidner04} and \citet{maschberger07}. In Sect.~\ref{subsect_igecmf} we describe the theory of the IGECMF, followed by an exemplification of how the IGECMF emerges from the ECMF for different star formation activities in Sect.~\ref{subsect_exempl}. The distribution function of SFRs, $F(\mathrm{SFR})$, and the SFH are compared in Sect.~\ref{subsect_sfh}.

\subsection{Improving optimal sampling} \label{subsect_sampling}

A simple and commonly used method of discretizing a parental distribution function is by random sampling. The distribution function is perceived as a probability distribution function from which values are diced using a generation function. Such an ensemble of sampled values naturally shows statistical deviations from the parental distribution function \citep{kroupa13rev}. However, for a wide range of applications it is necessary to accurately extract the number of objects as well as their individual masses from an arbitrary distribution function. This mass distribution function, denoted by $\xi (M)$, is described as

   \begin{equation} \label{xi}
      \xi (M) = \frac{\mathrm{d} N}{\mathrm{d} M} ,
   \end{equation}

\noindent where $\mathrm{d} N$ is the number of objects in the mass interval $M$ to $M + \mathrm{d} M$. The particular functional form of the distribution function, $\xi (M)$, is specified below and is not required for the general approach.

The outcome of any sampling technique should reproduce the number distribution, $\mathrm{d} N/\mathrm{d} M$, as well as the mass distribution, $M ~ \mathrm{d} N/\mathrm{d} M$, as precisely as possible. If this is the case for a sample of objects, the corresponding distribution function is called "fully populated" throughout this paper. Moreover, for any lower and upper mass limits, $M_{\mathrm{min}}$ and $M_{\mathrm{max}}$, respectively, the total number of objects, $N_{\mathrm{tot}}$, and the total mass, $M_{\mathrm{tot}}$, should agree with the analytical computation of $N_{\mathrm{tot}}$ as well as of $M_{\mathrm{tot}}$ at the same time:

   \begin{equation} \label{N_tot}
        N_{\mathrm{tot}} = \int^{M_{\mathrm{max}}}_{M_{\mathrm{min}}} { \xi (M) ~ \mathrm{d} M } ,
   \end{equation}

   \begin{equation} \label{M_tot}
        M_{\mathrm{tot}} = \int^{M_{\mathrm{max}}}_{M_{\mathrm{min}}} { M ~ \xi (M) ~ \mathrm{d} M } .
   \end{equation}

\noindent Thus, the quality of a sampling method can be measured by how accurately the outcome resembles $\mathrm{d} N/\mathrm{d} M$ and $M ~ \mathrm{d} N/\mathrm{d} M$ and how well the actual values for $N_{\mathrm{tot}}$ and $M_{\mathrm{tot}}$ agree with the analytical values (Eqs.~\ref{N_tot} and \ref{M_tot}). Compliance with one of these four conditions does not imply that one or all other conditions are fulfilled as well.

One technique used to do this is the optimal sampling method developed by \citet{kroupa13rev} and later incorporated into the extended software package originally published by \citet{pflamm-altenburg06}. It is designed to generate a population of stars from the initial mass function (IMF). The procedure requires the analytical form of the IMF, the physical upper mass limit for stars, $m_{\mathrm{max}}$, and the total stellar mass of the embedded SC, $M_{\mathrm{ecl}}$. As shown by \citet{kroupa13rev}, their Fig.~1, optimal sampling nicely reproduces the shape of the IMF, $\xi_{\mathrm{IMF}} (M) = \mathrm{d} N/\mathrm{d} M$, without introducing any Poisson noise. However, a closer look reveals that optimal sampling does not fulfill Eq.~\ref{N_tot}, as we show below.

Is it possible at all to devise a sampling technique that fulfills all four conditions and works without adding stochastic fluctuations to the outcome? It is, as we develop in the following.

Starting with Eqs.~\ref{N_tot} and \ref{M_tot}, we divide both integrals into $N_{\mathrm{tot}}$ separate integrals, each integral representing one individual object:

   \begin{align} \label{N_splitting}
   \begin{split}
        N_{\mathrm{tot}} & = \int^{m_{N_{\mathrm{tot}}}}_{M_{\mathrm{min}}} { \xi (M) ~ \mathrm{d} M } + \int^{m_{N_{\mathrm{tot}}-1}}_{m_{N_{\mathrm{tot}}}} { \xi (M) ~ \mathrm{d} M } + ... \: + \\ \int^{m_i}_{m_{i+1}} & { \xi (M) ~ \mathrm{d} M } + ... + \int^{m_2}_{m_3} { \xi (M) ~ \mathrm{d} M } + \int^{M_{\mathrm{max}}}_{m_2} { \xi (M) ~ \mathrm{d} M } ,
   \end{split}
   \end{align}

   \begin{align} \label{M_splitting}
   \begin{split}
        M_{\mathrm{tot}} & = \int^{m_{N_{\mathrm{tot}}}}_{M_{\mathrm{min}}} { M \xi (M) ~ \mathrm{d} M } + \int^{m_{N_{\mathrm{tot}}-1}}_{m_{N_{\mathrm{tot}}}} { M \xi (M) ~ \mathrm{d} M } + ... \: + \\ \int^{m_i}_{m_{i+1}} & { M \xi (M) ~ \mathrm{d} M } + ... + \int^{m_2}_{m_3} { M \xi (M) ~ \mathrm{d} M } + \int^{M_{\mathrm{max}}}_{m_2} { M \xi (M) ~ \mathrm{d} M } ,
   \end{split}
   \end{align}

\noindent since in total there are $N_{\mathrm{tot}}$ objects. With $m_1 = M_{\mathrm{max}}$, the index of the upper limit, $i$, of each separate integral enumerates the individual objects. Thus, each separate integral must fulfill the two following requirements:

\begin{enumerate}
\item Each integral must give one object. Integration of $\xi (M)$ within the limits $m_i$ and $m_{i+1}$ yields exactly unity:

   \begin{equation} \label{cond1}
      1 = \int^{m_i}_{m_{i+1}} { \xi (M) ~ \mathrm{d} M } .
   \end{equation}

\item Then the mass of this $i$-th object, $M_i$, is determined by 

   \begin{equation} \label{condM} 
      M_i = \int^{m_i}_{m_{i+1}} { M ~ \xi (M) ~ \mathrm{d} M } ,
   \end{equation}

where the limits $m_i$ and $m_{i+1}$ have to be equal to those in Eq.~\ref{cond1}.
\end{enumerate}

\noindent These two requirements ensure that the number distribution, $\mathrm{d} N/\mathrm{d} M$, and the mass distribution, $M ~ \mathrm{d} N/\mathrm{d} M$, are reproduced and that $N_{\mathrm{tot}}$ and $M_{\mathrm{tot}}$ agree with the analytical values. Since $m_{i+1} < m_i$ with increasing number $i$ the objects become less massive.

As our work considers the formation of SC distributions, our task is to generate an ideal population of SCs. However, the underlying concept is so general that it can be applied to any other type of object. For the sake of simplicity, we assumed that the mass distribution function of SCs follows a one-part power law with the index $\beta$

   \begin{equation} \label{xi_powerlaw}
      \xi (M) = k \left( \frac{M}{M_{\mathrm{max}}} \right)^{- \beta} \end{equation}

\noindent within the lower and upper mass limit, $M_{\mathrm{min}}$ and $M_{\mathrm{max}}$, respectively. $k$ is a normalization constant. Similarly to \citet{weidner04}, this function is normalized as follows:

   \begin{equation} \label{norm}
      1 = \int^{M_{\mathrm{trunc}}}_{M_{\mathrm{max}}} { \xi (M) ~ \mathrm{d} M } ,
   \end{equation}

\noindent with a truncation mass $M_{\mathrm{trunc}} = \infty$. This leads to a normalization constant, $k$, 

   \begin{equation} \label{norm_k}
      k = (\beta - 1) ~ M_{\mathrm{max}}^{-1} .
   \end{equation}

\noindent Here and in all following equations, $\beta > 1$ must be fulfilled. Otherwise, the antiderivatives of $\xi (M)$ and $M ~ \xi (M)$ cannot be computed. 

These ingredients enable computing the individual masses of the SCs. Equation~\ref{cond1} implies for the $(i+1)$-th integration limit of any of the separate integrals from Eq.~\ref{N_splitting}: 

   \begin{align} \label{1_iterate}
\begin{split}
      1 & = \int^{m_i}_{m_{i+1}} { \xi (M) ~ \mathrm{d} M } = M_{\mathrm{max}}^{\beta-1} \left( m_{i+1}^{1-\beta} - m_{i}^{1-\beta} \right) \\
      \Longleftrightarrow \quad m_{i+1} & = \left( m_{i}^{1-\beta} + M_{\mathrm{max}}^{1-\beta} \right)^{\frac{1}{1-\beta}} \quad \mathrm{ with } \quad m_1 = M_{\mathrm{max}} ,
\end{split}
   \end{align}

\noindent which allows iteratively determining the integration limits of all separate integrals in Eq.~\ref{N_splitting}. With these, the individual masses of all SCs of the ideal population can be computed, so that the $i$-th SC has a mass of

   \begin{align} \label{M_iterate}
\begin{split}
      M_i & = \int^{m_{i}}_{m_{i+1}} { M ~ \xi (M) ~ \mathrm{d} M } = \\ & =
  \begin{cases}
 M_{\mathrm{max}} ~ ( \ln m_{i} - \ln m_{i+1} )  &, \ \beta = 2 \\
 \frac{\beta - 1}{2 - \beta} ~ M_{\mathrm{max}}^{\beta - 1} ~ ( m_{i}^{2 - \beta} - m_{i+1}^{2 - \beta} )  &, \ \beta \neq 2 . \\
  \end{cases}
\end{split}
   \end{align}

\noindent Moreover, using Eq.~\ref{xi_powerlaw}, the expected total number of objects, $N_{\mathrm{tot}}$, and their total mass, $M_{\mathrm{tot}}$, as in Eqs.~\ref{N_tot} and \ref{M_tot}, can be evaluated analytically by replacing the lower and upper integration limits in Eqs.~\ref{1_iterate} and \ref{M_iterate} with $M_{\mathrm{min}}$ and $M_{\mathrm{max}}$, respectively: 

   \begin{equation} \label{N_analyt}
      N_{\mathrm{tot}} = \int^{M_{\mathrm{max}}}_{M_{\mathrm{min}}} { \xi (M) ~ \mathrm{d} M } = \left( \frac{M_{\mathrm{max}}}{M_{\mathrm{min}}} \right)^{\beta - 1} - 1 ,
   \end{equation}

   \begin{align} \label{M_analyt} 
\begin{split}
      M_{\mathrm{tot}} & = \int^{M_{\mathrm{max}}}_{M_{\mathrm{min}}} { M ~ \xi (M) ~ \mathrm{d} M } \\ & =
  \begin{cases}
 M_{\mathrm{max}} ~ \left( \ln M_{\mathrm{max}} - \ln M_{\mathrm{min}} \right)  &, \ \beta = 2 \\
 M_{\mathrm{max}} ~ \left[ \frac{\beta - 1}{2 - \beta} ~ \left( 1 - \left( \frac{ M_{\mathrm{min}} }{ M_{\mathrm{max}} } \right)^{2 - \beta} \right) \right]  &, \ \beta \neq 2 . \\
  \end{cases}
\end{split}
   \end{align}

We illustrate the performance of the introduced sampling technique and the comparison to the original optimal sampling method \citep{kroupa13rev} with an exemplary calculation. To quantify this comparison, we generated SCs with both sampling techniques based on the following assumptions:

\begin{itemize}
\item[\small$\bullet$] Mass distribution function, $\xi (M)$: The masses of the SCs are distributed according to Eq.~\ref{xi_powerlaw} with an index $\beta = 2.0$. For any other value for the index, one obtains the same qualitative results, therefore these results are omitted here.
\item[\small$\bullet$] Lower and upper limits: The lower limit for SCs is assumed to be constantly $M_{\mathrm{min}} = 5~\mathrm{M}_{\odot}$. The upper limit is varied in the range $10~\mathrm{M}_{\odot} < M_{\mathrm{max}} < 10^7~\mathrm{M}_{\odot}$.
\item[\small$\bullet$] Normalization: 
\begin{itemize}
\item[\tiny$\bullet$] New improved optimal sampling method: Using Eq.~\ref{norm} implies Eq.~\ref{norm_k} so that the normalization solely depends on $M_{\mathrm{max}}$ because $\beta$ is fixed.
\item[\tiny$\bullet$] Optimal sampling: The normalization is calculated as in \citet{pflamm-altenburg06} and requires the lower and upper limits, $M_{\mathrm{min}}$ and $M_{\mathrm{max}}$ and the total mass of the population, $M_{\mathrm{tot}}$. Thus, the analytical expectation for $M_{\mathrm{tot}}$ from Eq.~\ref{M_analyt} is provided as an input variable.
\end{itemize}
The normalizations of both sampling techniques lead to quantitatively the same $\xi (M)$ if $M_{\mathrm{min}}$, $M_{\mathrm{max}}$, and $M_{\mathrm{tot}}$ are equal for both techniques.
\end{itemize}

\begin{figure}[t]
\includegraphics[angle=-90, width=0.488\textwidth]{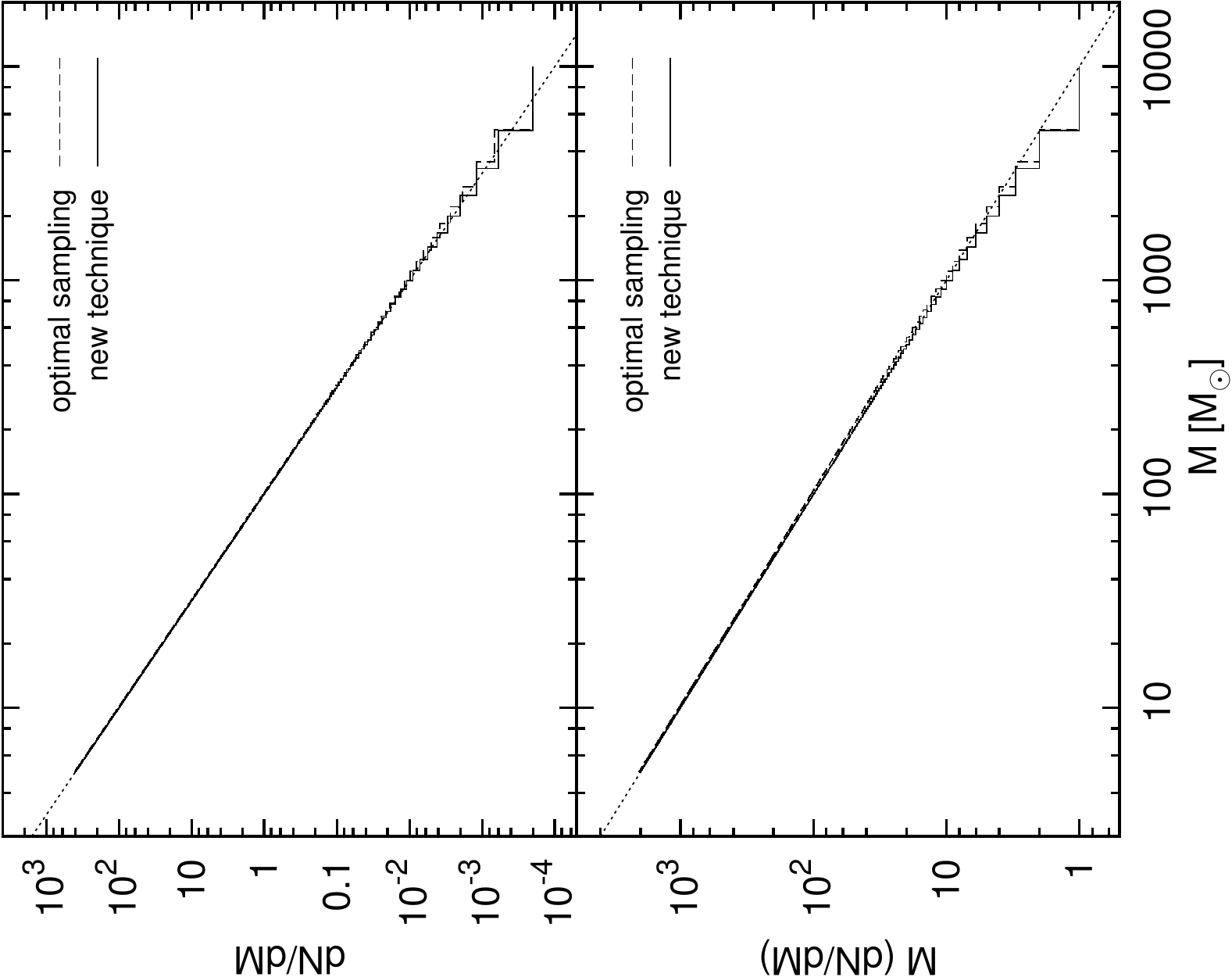}
\caption[Optimal sampling and the new sampling technique: Comparison of $\mathrm{d} N/\mathrm{d} M$ and $M ~ \mathrm{d} N/\mathrm{d} M$ with the expectations]{Comparison of the resulting number distribution, $\mathrm{d} N/\mathrm{d} M$, (upper panel) and mass distribution, $M ~ \mathrm{d} N/\mathrm{d} M$, (lower panel) obtained from optimal sampling (dashed lines) and the new sampling technique (continuous lines). The underlying mass distribution function, $\xi (M)$, is overplotted with a dotted line. Here, an exemplary case with $M_{\mathrm{max}} = 10$\,$000~\mathrm{M}_{\odot}$ is shown.}
\label{fig_xi_NM}
\end{figure}

Populations of SCs were generated for the above range of $M_{\mathrm{max}}$ using both sampling techniques. For an exemplary case with $M_{\mathrm{max}} = 10$\,$000~\mathrm{M}_{\odot}$, the distributions $\mathrm{d} N/\mathrm{d} M$ and $M ~ \mathrm{d} N/\mathrm{d} M$ are shown in Fig.~\ref{fig_xi_NM}. The result of optimal sampling is indicated with a dashed line, while a continuous line is used for the result of the new sampling technique. The parental distribution function, $\xi (M)$ and $M ~ \xi (M)$, is overplotted with a dotted line. Clearly, both sampling methods reproduce the parental function even if the resulting distributions differ slightly from each other. For lower or higher values of $M_{\mathrm{max}}$, these distributions are shifted to lower or higher masses, but look qualitatively the same. From this point of view, none of the sampling techniques can be favored over the other. 

\begin{figure}[t]
\includegraphics[angle=-90, width=0.488\textwidth]{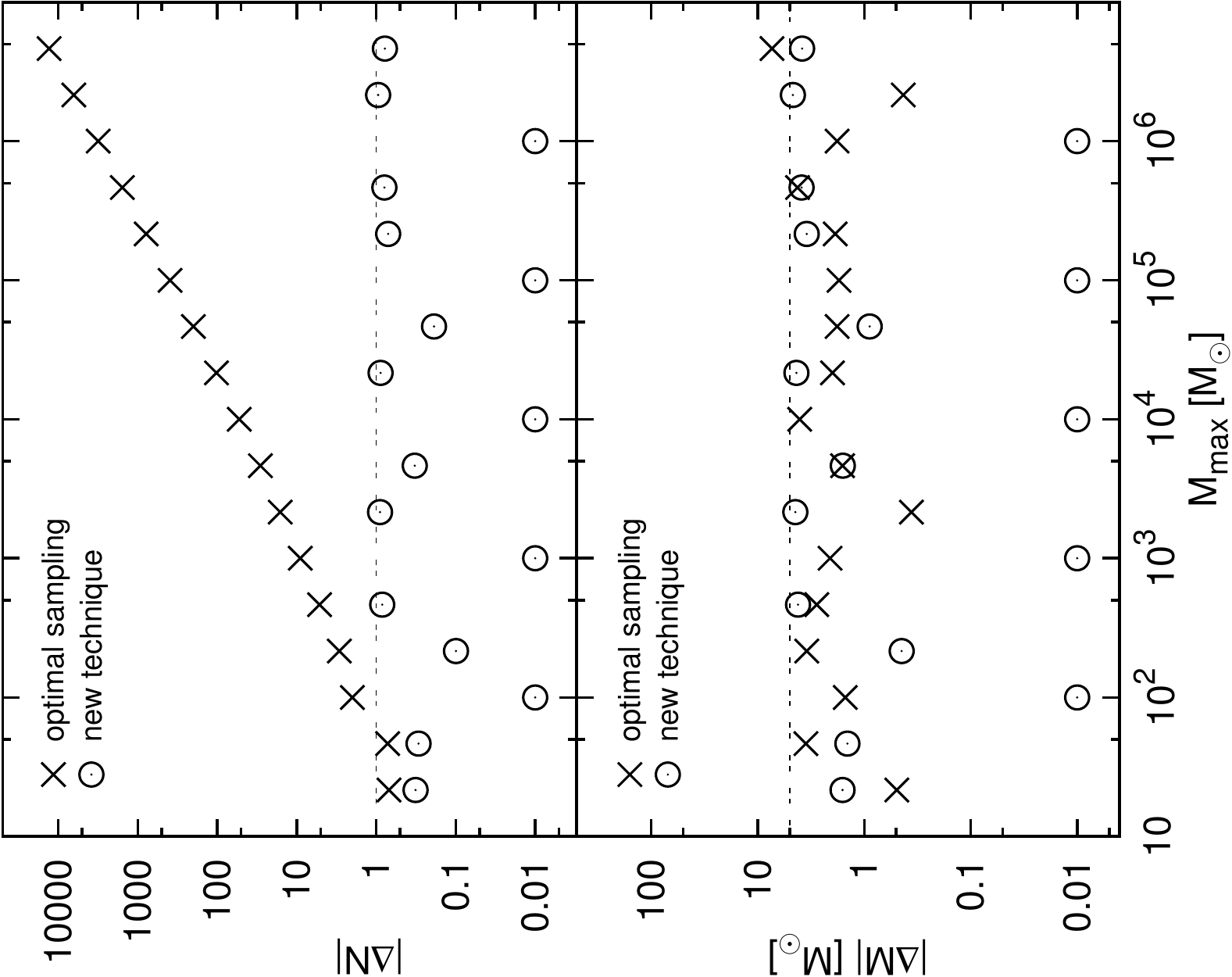}
\caption[Optimal sampling and the new sampling technique: Comparison of the deviations $|\Delta N|$ and $|\Delta M|$ from the expectations]{Comparison of the resulting difference, $|\Delta N|$ and $|\Delta M|$, obtained from optimal sampling (crosses) and the new sampling technique (open circles) with respect to the analytical expectation for $N_{\mathrm{tot}}$ and $M_{\mathrm{tot}}$ (Eqs.~\ref{N_analyt} and \ref{M_analyt}). The dashed lines indicate up to which level the deviations are acceptable ($\Delta N = 1$ and $\Delta M = 5~\mathrm{M}_{\odot}$). The corresponding values can be found in Table~\ref{tab_sampling}.}
\label{fig_Delta_NM}
\end{figure}

In addition, the resulting total numbers of SCs, $N_{\mathrm{tot}}$, and their total masses, $M_{\mathrm{tot}}$, of both sampling methods are compared to the analytical expectations from Eqs.~\ref{N_analyt} and \ref{M_analyt}. For each run we calculated the absolute deviations with respect to $N_{\mathrm{tot}}$, $|\Delta N|$, and to $M_{\mathrm{tot}}$, $|\Delta M|$. They can be found in Table~\ref{tab_sampling} together with the results. The absolute differences, $|\Delta N|$ and $|\Delta M|$, obtained from optimal sampling are marked with crosses, while those from the new sampling technique are drawn as open circles in Fig.~\ref{fig_Delta_NM}. The acceptable deviation is one SC of the lowest mass, that is,~$\Delta N = 1$ and $\Delta M = M_{\mathrm{min}} = 5~\mathrm{M}_{\odot}$ because it cannot be guaranteed that the expectation for $N_{\mathrm{tot}}$ (Eq.~\ref{N_analyt}) will give an exact integer value. These thresholds are indicated with dashed lines in Fig.~\ref{fig_Delta_NM}. Every data point that lies below this is compliant with the analytical expectation (Eqs.~\ref{N_analyt} and \ref{M_analyt}). For a certain combination of parameters ($\beta = 2.0$, $M_{\mathrm{min}} = 5~\mathrm{M}_{\odot}$, $M_{\mathrm{max}} = 10^x~\mathrm{M}_{\odot}$ with $x$ being a natural number), a natural number is analytically obtained for the total number of SCs, $N_{\mathrm{tot}}$ (Eq.~\ref{N_analyt}). The new sampling technique always leads to a discrete $N_{\mathrm{tot}}$ (cf. Table~\ref{tab_sampling}), so that for the upper combination of parameters it simply exactly reproduces the analytical expectation for $N_{\mathrm{tot}}$ and $M_{\mathrm{tot}}$. In these cases, the deviations $|\Delta N|$ or $|\Delta M|$ are thus zero and are plotted at 0.01 in Fig.~\ref{fig_Delta_NM} because otherwise these data points will be missing since we used a logarithmic scale.

\begin{table*}[t]
\caption[Results from optimal sampling and the new sampling technique]{Comparison of optimal sampling and the new sampling technique in terms of their absolute differences, $|\Delta N|$ and $|\Delta M|$, from the analytical expectations for $N_{\mathrm{tot}}$ and $M_{\mathrm{tot}}$ (Eqs.~\ref{N_analyt} and \ref{M_analyt}) as a function of the upper mass limit, $M_{\mathrm{max}}$. All deviations above the acceptable level ($\Delta N = 1$ and $\Delta M = 5~\mathrm{M}_{\odot}$) are shown in bold. All values are rounded.}
\label{tab_sampling}
\centering
\begin{tabular}{r|rr|rrrr|rrrr}
\hline \hline
 & \multicolumn{2}{c|}{analytical value} & \multicolumn{4}{c|}{optimal sampling} & \multicolumn{4}{c}{new sampling method} \\
\hline
$M_{\mathrm{max}}$ & $N_{\mathrm{tot}}$ & $M_{\mathrm{tot}}$ & $N_{\mathrm{tot}}$ & $|\Delta N|$ & $M_{\mathrm{tot}}$ & $|\Delta M|$ & $N_{\mathrm{tot}}$ & $|\Delta N|$ & $M_{\mathrm{tot}}$ & $|\Delta M|$ \\
$[\mathrm{M}_{\odot}]$ &  & $[\mathrm{M}_{\odot}]$ &  &  & $[\mathrm{M}_{\odot}]$ & $[\mathrm{M}_{\odot}]$ &  &  & $[\mathrm{M}_{\odot}]$ & $[\mathrm{M}_{\odot}]$ \\
\hline
21.5 & 3.3 & 31.5 & 4 & 0.7 & 31.0 & 0.5 & 3 & 0.3 & 29.9 & 1.6 \\
46.4 & 8.3 & 103.4 & 9 & 0.7 & 99.9 & 3.5 & 8 & 0.3 & 102.0 & 1.4 \\
100\hphantom{.0} & 19\hphantom{.0} & 299.6 & 21 & \textbf{2}\hphantom{.0} & 298.1 & 1.5 & 19 & 0\hphantom{.0} & 299.6 & 0.0 \\
215.4 & 42.1 & 810.8 & 45 & \textbf{2.9} & 807.3 & 3.5 & 42 & 0.1 & 810.3 & 0.4 \\
464.2 & 91.8 & 2$\,$103.0 & 97 & \textbf{5.2} & 2$\,$100.2 & 2.8 & 91 & 0.8 & 2$\,$098.8 & 4.2 \\
1$\,$000\hphantom{.0} & 199\hphantom{.0} & 5$\,$298.3 & 208 & \textbf{9}\hphantom{.0} & 5$\,$296.2 & 2.1 & 199 & 0\hphantom{.0} & 5$\,$298.3 & 0.0 \\
2$\,$154.4 & 429.9 & 13$\,$068.5 & 446 & \textbf{16.1} & 13$\,$068.1 & 0.4 & 429 & 0.9 & 13$\,$064.0 & 4.4 \\
4$\,$641.6 & 927.3 & 31$\,$717.7 & 956 & \textbf{28.7} & 31$\,$716.1 & 1.6 & 927 & 0.3 & 31$\,$716.1 & 1.6 \\
10$\,$000\hphantom{.0} & 1$\,$999\hphantom{.0} & 76$\,$009.0 & 2$\,$052 & \textbf{53}\hphantom{.0} & 76$\,$005.0 & 4.0 & 1$\,$999 & 0\hphantom{.0} & 76$\,$009.0 & 0.0 \\
21$\,$544.3 & 4$\,$307.9 & 180$\,$292.4 & 4$\,$410 & \textbf{102.1} & 180$\,$290.4 & 2.0 & 4$\,$307 & 0.9 & 180$\,$288.0 & 4.3 \\
46$\,$415.9 & 9$\,$282.2 & 424$\,$053.7 & 9$\,$481 & \textbf{198.8} & 424$\,$051.9 & 1.8 & 9$\,$282 & 0.2 & 424$\,$052.8 & 0.9 \\
100$\,$000\hphantom{.0} & 19$\,$999\hphantom{.0} & 990$\,$348.8 & 20$\,$391 & \textbf{392}\hphantom{.0} & 990$\,$347.0 & 1.7 & 19$\,$999 & 0\hphantom{.0} & 990$\,$348.8 & 0.0 \\
215$\,$443.5 & 43$\,$087.7 & 2$\,$299$\,$000.7 & 43$\,$868 & \textbf{780.3} & 2$\,$298$\,$998.8 & 1.9 & 43$\,$087 & 0.7 & 2$\,$298$\,$997.2 & 3.5 \\
464$\,$158.9 & 92$\,$830.8 & 5$\,$309$\,$301.9 & 94$\,$395 & \textbf{1$\,$564.2} & 5$\,$309$\,$297.7 & 4.2 & 92$\,$830 & 0.8 & 5$\,$309$\,$298.1 & 3.9 \\
1$\,$000$\,$000\hphantom{.0} & 199$\,$999\hphantom{.0} & 12$\,$206$\,$072.6 & 203$\,$154 & \textbf{3$\,$155}\hphantom{.0} & 12$\,$206$\,$070.8 & 1.8 & 199$\,$999 & 0\hphantom{.0} & 12$\,$206$\,$072.6 & 0.0 \\
2$\,$154$\,$434.7 & 430$\,$885.9 & 27$\,$950$\,$776.1 & 437$\,$276 & \textbf{6$\,$390.1} & 27$\,$950$\,$776.5 & 0.4 & 430$\,$885 & 0.9 & 27$\,$950$\,$771.4 & 4.7 \\
4$\,$641$\,$588.8 & 928$\,$316.8 & 63$\,$780$\,$672.7 & 941$\,$318 & \textbf{13$\,$001.2} & 63$\,$780$\,$665.3 & \textbf{7.4} & 928$\,$316 & 0.8 & 63$\,$780$\,$668.8 & 3.8 \\
\hline
\end{tabular}
\end{table*}

For $|\Delta M|,$ both approaches lead to a total mass, $M_{\mathrm{tot}}$, which agrees with the analytical expectation from Eq.~\ref{M_analyt} (lower panel of Fig.~\ref{fig_Delta_NM}, Table~\ref{tab_sampling}). However, we note that at the largest $M_{\mathrm{max}}$ the data point belonging to optimal sampling has $|\Delta M| = 7.4~\mathrm{M}_{\odot}$ (Table~\ref{tab_sampling}) and is therefore located slightly above the threshold in Fig.~\ref{fig_Delta_NM}. The reason for this is unclear, but it might be related to the accumulation of tiny random deviations.

For $|\Delta N|$, the upper panel of Fig.~\ref{fig_Delta_NM} shows that in all cases the total number of SCs generated by the new sampling technique agrees with the analytical expectation (Eq.~\ref{N_analyt}). However, this finding does not apply to optimal sampling: All runs with $M_{\mathrm{max}} \ge 100~\mathrm{M}_{\odot}$ produce more SCs than expected from Eq.~\ref{N_analyt}. As can be seen, the surplus number of SCs increase with increasing $M_{\mathrm{max}}$ (Table~\ref{tab_sampling}, upper panel of Fig.~\ref{fig_Delta_NM}). 

In conclusion, optimal sampling is not optimal in the sense that it does not lead to the correct total number of SCs, $N_{\mathrm{tot}}$, even if it reproduces the total mass, $M_{\mathrm{tot}}$, and the sampled SCs are distributed according to the parental distribution function (Fig.~\ref{fig_xi_NM}). In contrast, we introduce here for the first time a sampling method where the outcome is able to resemble the underlying distribution function in terms of $\mathrm{d} N/\mathrm{d} M$ and $M ~ \mathrm{d} N/\mathrm{d} M$ and has a total number of SCs, $N_{\mathrm{tot}}$, and a total mass, $M_{\mathrm{tot}}$, which fully agrees with the analytical expectations (Eqs.~\ref{N_analyt} and \ref{M_analyt}). Thus, the introduced new sampling method can be termed improved optimal sampling.

We add that the normalization (Eq.~\ref{norm}) may seem to be arbitrary since there is no physical reason why the integration of a distribution function between the theoretical upper mass limit, $M_{\mathrm{max}}$, and the truncation mass, $M_{\mathrm{trunc}} = \infty$, should exactly yield unity. However, this mathematical step is essential since it allows $M_{\mathrm{max}}$ to vary: Here, $M_{\mathrm{max}}$ cannot be treated as a constant quantity since it depends on the SFR, as we derive in Sect.~\ref{subsect_ecmf}. For instance, for a stellar IMF, the most massive star, $m_{\mathrm{max}}$, is determined by the total stellar mass of the SC, $M_{\mathrm{ecl}}$, according to the $m_{\mathrm{max}}-M_{\mathrm{ecl}}$ relation \citep{weidner_k_b10, weidner_k_pa13}, while the stellar truncation mass would be $M_{\mathrm{trunc}} = 150~\mathrm{M}_{\odot}$ \citep[e.g.,~][]{weidner_kroupa04}. This already shows the universal applicability of the new sampling technique: For stars, the underlying distribution function would be a multiple-part power-law IMF to which the new sampling technique can be applied in the same manner as for SCs with an ECMF defined by a one-part power law. More generally, this method can be used to sample any type of object that can be represented by a distribution function that is a piecewise power law and where avoidance of Poisson noise is desired.

The following instruction describes how a population of objects can be created with the new sampling technique:

\begin{enumerate}
\item Define the (mass) distribution function, $\xi (M)$, and its lower and upper limits, $M_{\mathrm{min}}$ and $M_{\mathrm{max}}$. Remember that for SCs $M_{\mathrm{max}}$ depends on the SFR (Sect.~\ref{subsect_ecmf}), while for stars the most massive one, $m_{\mathrm{max}}$, is determined by the total stellar mass of the SC, $M_{\mathrm{ecl}}$, according to the $m_{\mathrm{max}}-M_{\mathrm{ecl}}$ relation.
\item Apply a normalization. Here, we assumed Eq.~\ref{norm}, but other normalizations are conceivable as well. For stars, the assumption of a certain $M_{\mathrm{ecl}}$ directly normalizes $\xi (M)$.
\item Determine the integration limits of Eq.~\ref{N_splitting} in consideration of Eq.~\ref{cond1}. If the antiderivative of Eq.~\ref{cond1} cannot be calculated analytically, the integration limits can also be computed numerically by starting at $M_{\mathrm{max}}$ and integrating downward. If a one-part power law with index $\beta$ is assumed and the same normalization is used (Eqs.~\ref{norm} and \ref{norm_k}), all integration limits can be directly calculated from Eq.~\ref{1_iterate}.
\item Using the integration limits, compute the individual masses of all the objects from Eq.~\ref{condM}. Again, this can be done numerically if the respective antiderivative does not exist. Using the same assumptions as we did here, the individual masses can be directly calculated from Eq.~\ref{M_iterate}.
\end{enumerate}

The advantages of the new improved optimal sampling technique are that it fulfills all of the four conditions stated at the beginning of this section, meaning that the number distribution and the mass distribution match the underlying mass distribution function. In addition, the total number of objects, $N_{\mathrm{tot}}$, and their total mass, $M_{\mathrm{tot}}$, match analytical expectations (Eqs.~\ref{N_tot} and \ref{M_tot}). As requested, the outcome is free of stochastic fluctuations. Moreover, if the antiderivatives of $\mathrm{d} N/\mathrm{d} M$ and $M ~ \mathrm{d} N/\mathrm{d} M$ (cf.~Eqs.~\ref{N_tot} - \ref{condM}) can be calculated analytically, then the whole sampling can be performed analytically as well, which saves computational time.

\subsection{Embedded cluster mass function (ECMF)} \label{subsect_ecmf}
The ECMF is the mass distribution function of young, embedded SCs that were formed during one star cluster formation epoch (SCFE). Observations suggest that the stellar masses of young SCs are distributed according to a power law with index $\beta$: 

   \begin{equation} \label{ecmf}
      \xi_{\mathrm{ECMF}} (M) = \frac{\mathrm{d} N_{\mathrm{ECMF}}}{\mathrm{d} M} = k \left( \frac{M}{M_{\mathrm{max}}} \right)^{- \beta} .
   \end{equation}

\noindent $M_{\mathrm{max}}$ is the stellar upper mass limit for SCs formed during one SC formation epoch, $k$ a normalization constant, and $- \beta$ the slope of the ECMF lying in the range $1.6 \lesssim \beta \lesssim 2.5$ (direct measurements: \citet{zhang_fall99, bik03, degrijs03, hunter03, lada_lada03, fall04asp, gieles06, degrijs_anders06, mccrady_graham07, degrijs_goodwin08, dowell08, whitmore10, chandar10, chandar11}; derived from models: e.g., \citet{kroupa_boily02, weidner04}; see also \citet{degrijs03}, their Table~2, for slopes of the cluster luminosity function for different galaxies).

It is debated whether the ECMF is a pure power law \citep{whitmore07, whitmore10, chandar10, chandar11} or has a fundamental upper limit like a cutoff or an exponential turn-over at the high-mass end, which can be described by a Schechter function \citep{gieles06, gieles06letter, bastian08, larsen09, bastian12, bastian12rn}. A differentiation between the two types is very difficult because of the low number of high-mass SCs \citep[e.g.,][]{bastian08, bastian12}. If the ECMF is indeed truncated, \citet{haas_anders10} did not expect the precise shape at the high-mass end to be important. They investigated how the choice of the sampling technique and the index of the ECMF alters the integrated galactic initial mass function (IGIMF) -- the analogon of the IGECMF for stars instead of SCs. They pointed out that an exponential turn-down and a truncation of the cluster mass function will have a similar effect on the IGIMF, for which reason the precise shape of the ECMF is not expected to be important. 

\citet{bonatto12}, for instance, simulated how a Schechter-type initial cluster mass function of galactic GCs evolves due to stellar evolution and dynamical mass-loss processes into a present-day mass function for different $M/L$ dependences on luminosity. Interestingly, the most realistic results were obtained for $M/L$ ratios increasing with luminosity -- as is observed for GCs -- with a truncation mass of $M_{\mathrm{trunc}} \approx 10^{10}~\mathrm{M}_{\odot}$, which means, effectively a pure power law without an upper limit. On the other hand, there might exist an upper mass limit for SCs since they form out of giant molecular clouds (GMCs) whose mass function is truncated at the high-mass end at least in M33, as reported by \citet{rosolowsky07}.

Since a completely limitless ECMF is unphysical, it is assumed that there is a theoretical upper mass limit for SCs, $M_{\mathrm{max}}$, which is not a fixed value, but depends on the SFR, as we derive below. Following \citet{weidner04}, we take a lower mass limit for newly born SCs of $M_{\mathrm{min}} = 5~\mathrm{M}_{\odot}$. The ECMF (Eq.~\ref{ecmf}) was assumed to be a pure power law ranging from $M_{\mathrm{min}}$ to the cutoff mass, $M_{\mathrm{max}}$, beyond which SCs cannot be formed. All following derivations are based on the findings from Sect.~\ref{subsect_sampling}, meaning that the same normalization and the new sampling method were used so that all results obtained there are applicable here. The choice of a deterministic sampling technique is motivated by \citet{pflamm-altenburg13} and \citet[see also references therein]{kroupa15review} since a self-regulated rather than a probabilistic or stochastic description of the emergence of an SC population out of a dense molecular cloud is consistent with the data. Thus, the total number of young SCs, $N_{\mathrm{ECMF}}$, of one SC formation epoch is given by Eq.~\ref{N_analyt}, 

   \begin{equation} \label{N_ecmf}
      N_{\mathrm{ECMF}} = \int^{M_{\mathrm{max}}}_{M_{\mathrm{min}}} { \xi_{\mathrm{ECMF}} (M) ~ \mathrm{d} M } = \left( \frac{M_{\mathrm{max}}}{M_{\mathrm{min}}} \right)^{\beta - 1} - 1 .
   \end{equation}

\noindent According to Eq.~\ref{M_analyt}, the total stellar mass of a young, embedded SC population, $M_{\mathrm{ECMF}}$, formed during one SC formation epoch, is determined by

   \begin{align} \label{M_ecmf} 
\begin{split}
      M_{\mathrm{ECMF}} & = \int^{M_{\mathrm{max}}}_{M_{\mathrm{min}}} { M ~ \xi_{\mathrm{ECMF}} (M) ~ \mathrm{d} M } \\ & =
  \begin{cases}
 M_{\mathrm{max}} ~ \left( \ln M_{\mathrm{max}} - \ln M_{\mathrm{min}} \right)  &, \ \beta = 2 \\
 M_{\mathrm{max}} ~ \left[ \frac{\beta - 1}{2 - \beta} ~ \left( 1 - \left( \frac{ M_{\mathrm{min}} }{ M_{\mathrm{max}} } \right)^{2 - \beta} \right) \right]  &, \ \beta \neq 2 . \\
  \end{cases}
\end{split}
   \end{align}

For all following computations we assumed the following for the SC formation process:

\begin{enumerate}
\item During one SC formation epoch, all SCs and the stars therein form coevally and represent a single-age SC population. The SC masses of this young SC population are always distributed according to the ECMF (Eq.~\ref{ecmf}) within the limits $M_{\mathrm{min}}$ and $M_{\mathrm{max}}$, implying that the ECMF is fully (or 'optimally') populated (Sect.~\ref{subsect_sampling}).
\item The index $\beta$ of the ECMF does not change with time.
\item An SC formation epoch is of duration $\delta t$, which is not a function of time.
\item During an SC formation epoch, the total mass of the young SC population, $M_{\mathrm{ECMF}}$, is formed at a constant SFR:

   \begin{equation} \label{mtotsfrdt}
      M_{\mathrm{ECMF}} = \mathrm{SFR} \cdot \delta t .
   \end{equation}

\end{enumerate}

\noindent The total mass of one SC population, $M_{\mathrm{ECMF}}$, can be calculated from Eq.~\ref{M_ecmf} if the lower and upper limit of the SC masses are known. Moreover, knowledge about $M_{\mathrm{ECMF}}$ and $\delta t$ allows extracting the underlying SFR: Rearranging Eq.~\ref{mtotsfrdt} using Eq.~\ref{M_ecmf} leads to an SFR of

\begin{align} \label{sfr}
 \mathrm{SFR} &= 
  \begin{cases}
 \frac{M_{\mathrm{max}}}{\delta t} ~ \left( \ln M_{\mathrm{max}} - \ln M_{\mathrm{min}} \right)  &, \ \beta = 2 \\
 \frac{ M_{\mathrm{max}} }{\delta t} ~ \frac{\beta - 1}{2 - \beta} ~ \left( 1 - \left( \frac{ M_{\mathrm{min}} }{ M_{\mathrm{max}} } \right)^{2 - \beta} \right)  &, \ \beta \neq 2 . \\
  \end{cases}
\end{align}

\noindent Since $M_{\mathrm{min}}$, $\beta$, and $\delta t$ are treated as constant quantities, the SFR (Eq.~\ref{sfr}) is determined by $M_{\mathrm{max}}$ alone. Because the ECMF is a function of $M_{\mathrm{max}}$ and $M_{\mathrm{max}}$ is correlated with the SFR, the ECMF implicitly depends on the SFR:

   \begin{equation} \label{ecmf_limits}
      \xi_{\mathrm{ECMF}} (M) \equiv \xi_{\mathrm{ECMF,SFR}} ( M_{\mathrm{min}} \le M \le M_{\mathrm{max}} (\mathrm{SFR}) ) .
   \end{equation}

\noindent Regrettably, the theoretical upper mass limit for SCs of a particular SC formation epoch, $M_{\mathrm{max}}$, is very hard to determine. However, the mass of the most massive SC of the same SC formation epoch, $M_{\mathrm{ecl,max}}$, can be estimated. The ansatz of the new sampling technique enables relating the theoretical upper mass limit, $M_{\mathrm{max}}$, and the mass of the heaviest SC, $M_{\mathrm{ecl,max}}$, to each other. In the first condition (Eq.~\ref{cond1}), $i = 1$ is assigned to the most massive SC since the SCs become less massive with increasing $i$ (Sect.~\ref{subsect_sampling}), so $m_i = m_1 = M_{\mathrm{max}}$

   \begin{equation} \label{1_eclmax}
      1 = \int^{M_{\mathrm{max}}}_{m_2} { \xi_{\mathrm{ECMF}} (M) ~ \mathrm{d} M } = \left( \frac{m_2}{M_{\mathrm{max}}} \right)^{1-\beta} - 1 .
   \end{equation}

\noindent Solving for $m_2$ gives

   \begin{equation} \label{M_low}
      m_2 = 2^{\frac{1}{1-\beta}} M_{\mathrm{max}} .
   \end{equation}

\noindent According to the second condition (Eq.~\ref{condM}), the mass of the most massive SC, $M_{\mathrm{ecl,max}}$, is determined by the integration limits from Eq.~\ref{1_eclmax} and replacing $m_2$ with Eq.~\ref{M_low} results in

   \begin{align} \label{M_eclmax}
\begin{split}
      M_{\mathrm{ecl,max}} & = \int^{M_{\mathrm{max}}}_{m_2} { M ~ \xi_{\mathrm{ECMF}} (M) ~ \mathrm{d} M } \\ & = 
  \begin{cases}
  ~ \left( \ln 2 \right) ~ M_{\mathrm{max}}  &, \ \beta = 2 \\
  ~ \frac{\beta - 1}{2 - \beta} ~ \left( 1 - 2^{\frac{2 - \beta}{1 - \beta}} \right) ~ M_{\mathrm{max}}  &, \ \beta \neq 2 . \\
  \end{cases}
\end{split}
   \end{align}

\noindent Inversely, the upper mass limit for SCs of one SC formation epoch, $M_{\mathrm{max}}$, as a function of the observed most massive SC, $M_{\mathrm{ecl,max}}$, reads

 \begin{align} \label{relation_max_eclmax} 
   M_{\mathrm{max}} &=
\begin{cases} 
  ~ \left( \ln 2 \right)^{-1} ~ M_{\mathrm{ecl,max}}  &, \ \beta = 2 \\
  ~ \frac{2 - \beta}{\beta - 1} ~ \left( 1 - 2^{\frac{2 - \beta}{1 - \beta}} \right)^{-1} ~ M_{\mathrm{ecl,max}}  &, \ \beta \neq 2 , \\
\end{cases}
   \end{align}

\noindent which allows relating $M_{\mathrm{ecl,max}}$ and SFR to each other by replacing $M_{\mathrm{max}}$ in Eq.~\ref{sfr} with Eq.~\ref{relation_max_eclmax}. From this arises the so-called SFR-$M_{\mathrm{ecl,max}}$ relation:

\begin{align} \label{sfr_M_eclmax}
 \mathrm{SFR} &= 
  \begin{cases}
 \frac{M_{\mathrm{ecl,max}}}{\delta t \cdot \ln 2} \left( \ln \left( \frac{M_{\mathrm{ecl,max}}}{\ln 2} \right) - \ln M_{\mathrm{min}} \right) &, \beta = 2 \\
 \frac{ M_{\mathrm{ecl,max}} }{\delta t} \left( 1 - 2^{\frac{2 - \beta}{1 - \beta}} \right)^{-1} \left( 1 - \left( \frac{\beta - 1}{2 - \beta} ~ \left( 1 - 2^{\frac{2 - \beta}{1 - \beta}} \right) \frac{ M_{\mathrm{min}} }{ M_{\mathrm{ecl,max}} } \right)^{2 - \beta} \right) \!\!\!\!\!\! &, \beta \neq 2 . \\
  \end{cases}
\end{align}

\noindent Indeed, observations suggest that $M_{\mathrm{ecl,max}}$ scales with SFR (see Fig.~\ref{fig_sfr_Mmax5} below) as found for example by~\citet{weidner04}. In Sect.~\ref{sect_dt} we analyze this SFR-$M_{\mathrm{ecl,max}}$ relation to determine the length of one SC formation epoch, $\delta t$.

\subsection{Integrated galactic embedded cluster mass function (IGECMF)} \label{subsect_igecmf}

The purpose of this contribution is to devise a mass distribution function describing how the birth stellar masses of young SCs are distributed after a formation episode that is much longer than one single SC formation epoch of length $\delta t$. In this respect, it does not matter whether SC formation takes place continuously or in bursty phases. We assumed that any SC formation episode can be divided into a certain number of SC formation epochs of length $\delta t$ and that all properties mentioned in Sect.~\ref{subsect_ecmf} apply to each epoch. Since the masses of SCs that formed during one SC formation epoch are distributed according to the ECMF, the superposition of all involved ECMFs will lead to the requested mass distribution function. Analogously to the ECMF, this time-integrated function is called the IGECMF. The IGECMF reveals how the birth stellar masses of SCs are distributed after a certain SC formation episode, but it does not take into account any changes of the individual SC masses afterward.

The ECMF of each SC formation epoch is determined by an individual $M_{\mathrm{max}}$ (cf.~Eqs.~\ref{ecmf} and \ref{norm_k}). Since $M_{\mathrm{max}}$ is dependent on the SFR (Eq.~\ref{sfr}, see also Eq.~\ref{ecmf_limits}), the distribution function of SFRs, called $F(\mathrm{SFR})$, is needed to obtain the IGECMF. It describes the number of SC formation epochs (SCFEs) $\mathrm{d} N_{\mathrm{SCFE}} (\mathrm{SFR})$ per SFR interval:

   \begin{equation} \label{F}
      F(\mathrm{SFR}) = \frac{\mathrm{d} N_{\mathrm{SCFE}} (\mathrm{SFR}) }{\mathrm{d} \mathrm{SFR}} .
   \end{equation}

The IGECMF will arise from the integration of the ECMF over the whole range of SFRs in which the ECMF (Eq.~\ref{ecmf}, see also Eq.~\ref{ecmf_limits}) is modulated by $F(\mathrm{SFR})$:

   \begin{equation} \label{igecmf}
      \xi_{\mathrm{IGECMF}} (M) = \int_{\mathrm{SFR_{min}}}^{\mathrm{SFR_{max}}} { \xi_{\mathrm{ECMF,SFR}} (M) ~ F(\mathrm{SFR}) ~ \mathrm{d} \mathrm{SFR} } .
   \end{equation}

\noindent The resulting IGECMF will have a unique shape because $F(\mathrm{SFR})$ carries information about the formation history, which is unique for any galaxy or galaxy cluster. Since the ECMF implicitly depends on SFR (Eq.~\ref{ecmf_limits}) and the inverse function $M_{\mathrm{max}}(\mathrm{SFR})$ cannot be calculated analytically (cf.~Eq.~\ref{sfr}), the integration of Eq.~\ref{igecmf} cannot be performed directly. For this reason, $M_{\mathrm{max}}$ is substituted for the integration variable SFR so that Eq.~\ref{igecmf} becomes

   \begin{equation} \label{igecmf_mod} 
      \xi_{\mathrm{IGECMF}} (M) = \int_{M_{\mathrm{max}}^{\mathrm{low}}}^{M_{\mathrm{max}}^{\mathrm{up}}} \xi_{\mathrm{ECMF}} (M, M_{\mathrm{max}}) ~ F(M_{\mathrm{max}}) ~ \frac{\mathrm{d} \mathrm{SFR}}{\mathrm{d} M_{\mathrm{max}}} ~ \mathrm{d} M_{\mathrm{max}} ,
   \end{equation}

\noindent with $M_{\mathrm{max}}^{\mathrm{low}}$ and $M_{\mathrm{max}}^{\mathrm{up}}$ being the new limits of the integration over all possible $M_{\mathrm{max}}$ corresponding to the lowest and highest SFRs, $\mathrm{SFR_{min}}$ and $\mathrm{SFR_{max}}$. The transformation from $F(\mathrm{SFR})$ to $F(M_{\mathrm{max}})$ is carried out in Sect.~\ref{sect_igecmf}. The derivatives of Eq.~\ref{sfr} are

   \begin{align}
        \frac{\mathrm{d} \mathrm{SFR}}{\mathrm{d} M_{\mathrm{max}}} &= 
  \begin{cases}
 \frac{1}{\delta t} ~ \left( 1 + \ln M_{\mathrm{max}} - \ln M_{\mathrm{min}} \right)  &, \ \beta = 2 \\
 \frac{1}{\delta t} ~ \left[ \frac{\beta - 1}{2 - \beta} ~ \left( 1 - (\beta - 1) \left( \frac{ M_{\mathrm{min}} }{ M_{\mathrm{max}} } \right)^{2 - \beta} \right) \right]  &, \ \beta \neq 2 . \\
  \end{cases}
   \end{align}

\noindent The integral in Eq.~\ref{igecmf_mod} is equivalent to a summation of ECMFs up to their individual $M_{\mathrm{max}}$. $F(M_{\mathrm{max}})$ determines how often each ECMF contributes to the overall IGECMF since it reveals how often the corresponding SFR occurred. Thus, this methodology is called the superposition principle.

It would be convenient to directly extract the total number of SCs, $N_{\mathrm{IGECMF}}$, as well as their total mass, $M_{\mathrm{IGECMF}}$, from the IGECMF. In the same way as for the ECMF, we use the new sampling technique by applying the criteria in Eqs.~\ref{cond1} and \ref{condM} to the IGECMF: The $i$-th SC is obtained from the IGECMF (Eq.~\ref{igecmf_mod}) if the integrations limits $m_{i+1}$ and $m_i$ are chosen such that the integral over the IGECMF is exactly unity:

   \begin{equation} \label{1_igecmf}
      1 = \int^{m_i}_{m_{i+1}} { \xi_{\mathrm{IGECMF}} (M) ~ \mathrm{d} M } .
   \end{equation}

\noindent Then the mass of the $i$-th SC is determined by 

   \begin{equation} \label{M_ecl_igecmf} 
      M_{\mathrm{ecl},i} = \int^{m_i}_{m_{i+1}} { M ~ \xi_{\mathrm{IGECMF}} (M) ~ \mathrm{d} M } ,
   \end{equation}

\noindent where the limits $m_{i+1}$ and $m_i$ must be equal to those in Eq.~\ref{1_igecmf}. Consequently, the total number of SCs, $N_{\mathrm{IGECMF}}$, with masses within the limits $M_{\mathrm{min}}$ and $M_{\mathrm{max}}$ resulting from a SC formation episode of length $\delta t$ is given by

   \begin{equation} \label{N_igecmf}
      N_{\mathrm{IGECMF}} = \int_{M_{\mathrm{min}}}^{M_{\mathrm{max}}} { \xi_{\mathrm{IGECMF}} (M) ~ \mathrm{d} M } ,
   \end{equation}

\noindent and its total mass, $M_{\mathrm{IGECMF}}$, can be calculated from

   \begin{equation} \label{M_igecmf}
      M_{\mathrm{IGECMF}} = \int^{M_{\mathrm{max}}}_{M_{\mathrm{min}}} { M ~ \xi_{\mathrm{IGECMF}} (M) ~ \mathrm{d} M } .
   \end{equation}

We focus on how the birth stellar masses of all SCs ever formed will be distributed after an SC formation episode of arbitrary duration. This mass distribution is computed by purely superposing the single-age SC populations of many SC formation events, allowing the SFR to change with time. Thus, neither stellar nor dynamical evolution leading to mass loss or even to the destruction of SCs are taken into account here. However, these effects and the impact of the tidal field are discussed in Sect.~\ref{sect_obs} and must be accounted for as soon as the derived SC mass distributions are compared to observed mass distributions of SCs.

\subsection{From the ECMF to the IGECMF -- an exemplification} \label{subsect_exempl}
\begin{figure*}[t]
\includegraphics[angle=-90, width=\textwidth]{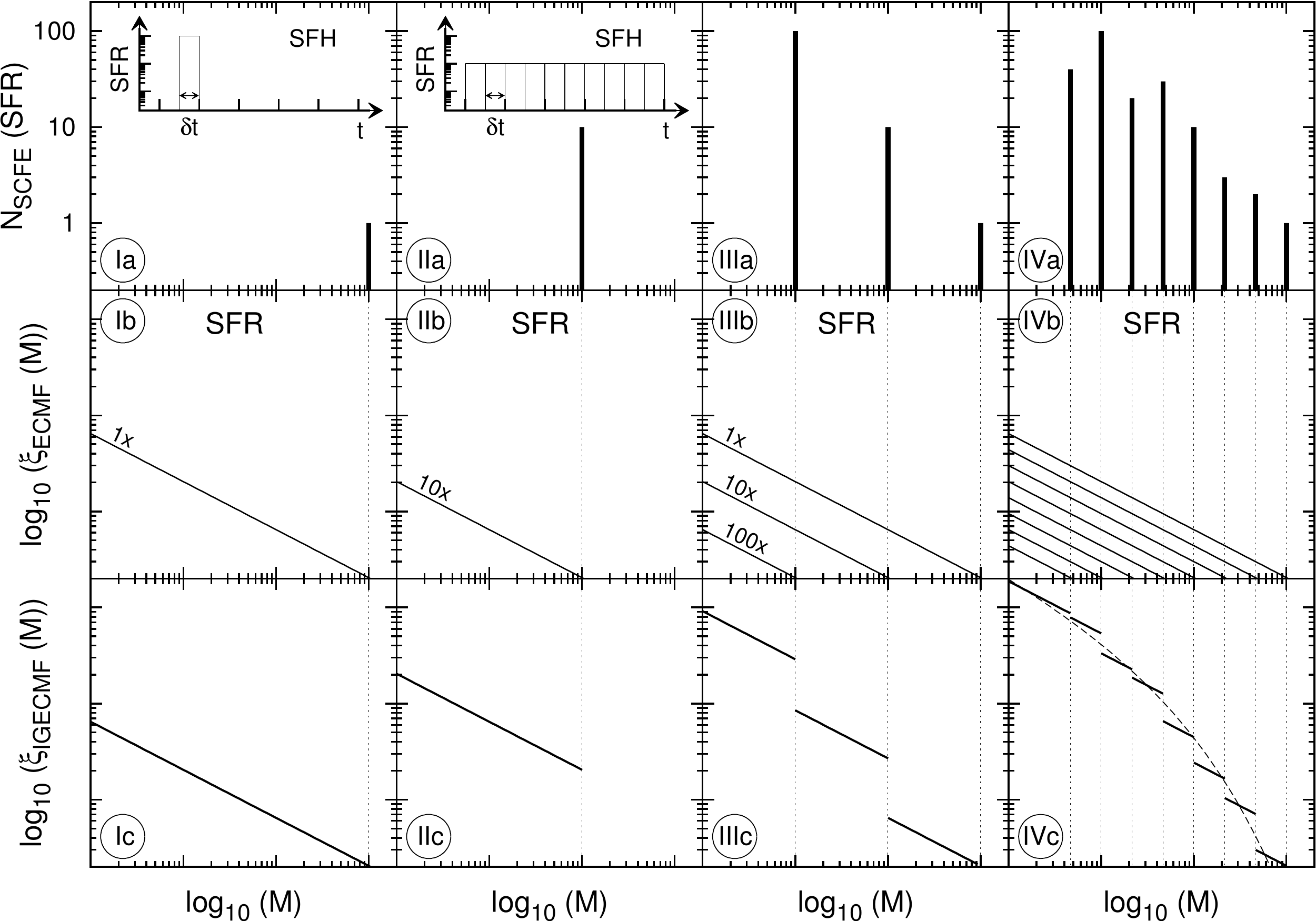}
\caption[Sketch: How star formation activities shape the IGECMF]{Sketch of how different star formation activities (top panels, 'a') influence the ECMFs (middle panels, 'b') and thereby shape the IGECMF (bottom panels, 'c') for four exemplary cases (columns I -- IV) according to the superposition principle.}
\label{fig_intro_plot}
\end{figure*}

Since the subject matter of the previous sections is very theoretical, the interrelation between the ECMF and the IGECMF is exemplified with Fig.~\ref{fig_intro_plot} in this section. We sketch how different star formation activities (top panels, labeled 'a') influence the ECMFs (middle panels, labeled 'b') and thereby shape the IGECMF (bottom panels, labeled 'c'). A double-logarithmic scale is used so that the power-law ECMFs appear as straight lines.

Starting with the easiest case (I), there is exactly one SC formation epoch of length $\delta t$ at a relative high SFR (Ia). Converting this SFR to its respective $M_{\mathrm{max}}$ (vertical thin dotted line in panel Ib) visualizes up to which mass the corresponding ECMF is populated once -- since only one formation epoch occurred (Ib). Thus, the respective IGECMF (Ic) is equal to the ECMF (Ib) since just one ECMF contributed.

In the second case (II), the SFR remains constant at a medium level over ten SC formation epochs, each of duration $\delta t$ (IIa). Again, converting the SFR to its respective $M_{\mathrm{max}}$ defines the ECMF (IIb), which is populated ten times in total. Thus, the corresponding IGECMF (IIc) is shifted upward by a factor of ten and is truncated sharply at $M_{\mathrm{max}}$, which is lower than in the first case (I).

From examples (I) and (II) it becomes clear that, on the one hand, the level of the SFR is relevant since it determines the upper mass limit of the ECMF, and on the other hand, that how often this particular SFR appeared constitutes how often the corresponding ECMF contributes to the IGECMF. This is exemplified in the third case (III), where SC formation takes place at different SFRs and each SFR appears a certain number of times (IIIa). Each SFR has its own respective ECMF, which is populated the number of times the SFR occurred (IIIb): The ECMF corresponding to the lowest SFR is populated one hundred times, while the ECMF belonging to the highest SFR is populated just once. According to the superposition principle, the IGECMF (IIIc) is composed of all contributing ECMFs. The summation has to be carried out in the mass ranges separated by the thin dotted lines by taking into account how often each ECMF occurred: In the highest mass range, only the ECMF corresponding to the highest SFR has to be considered, while in the lowest mass range all ECMFs contribute. Thus, one observes jumps in the IGECMF (IIIc).

The most realistic case is presented in the last column (IV). As in (IIIa), SC formation takes place at different SFRs and a different number of times (IVa), which defines how often each corresponding ECMF will be populated (IVb). Superposing all these ECMFs -- each multiplied by the number of formation epochs of the respective SFR -- leads to the IGECMF (IVc). Since the summation has to be performed separately in each mass range, the IGECMF exhibits jumps, as in the case before (IIIc).

If even more SC formation epochs occur at SFRs lying between the considered ones (IVa), even more but smaller jumps will appear in the IGECMF. It will develop a curved shape, as indicated by the dashed line (IVc), meaning that it becomes steeper toward the high-mass end.

\subsection[Comparing $F(\mathrm{SFR})$ and the star formation history (SFH)]{Comparing $F(\mathrm{SFR})$ and the star formation history (SFH)} \label{subsect_sfh}
The distribution function of SFRs, $F(\mathrm{SFR})$, represents all contributing star formation activities, but not their chronological order. Consequently, it is different from the SFH, which describes how the SFR evolves with time, meaning $\mathrm{SFH} \equiv \mathrm{SFR} (t)$, and should not be confused with it. However, $F(\mathrm{SFR})$ can be derived from a known SFH: The SFH has to be divided into single SC formation epochs of length $\delta t$ and for each SC formation epoch the associated SFR has to be determined. Counting how often ($N_{\mathrm{SCFE}}$) every SFR appeared reveals the distribution of SFRs, $N_{\mathrm{SCFE}} (\mathrm{SFR})$, from which $F(\mathrm{SFR})$ can be calculated by applying Eq.~\ref{F}. We show this in detail for six considered SFHs in Sect.~\ref{sect_F}.

For instance, a constant SFH at a level of $\mathrm{SFR} = 10~\mathrm{M}_{\odot}\mathrm{yr}^{-1}$ continuing over ten SC formation epochs leads to $F(\mathrm{SFR})$ being a delta function at $\mathrm{SFR} = 10~\mathrm{M}_{\odot}\mathrm{yr}^{-1}$ normalized to 10. For comparison, a star burst lasting over one SC formation epoch at $\mathrm{SFR} = 100~\mathrm{M}_{\odot}\mathrm{yr}^{-1}$ corresponds to $F(\mathrm{SFR})$ being a delta function as well, but at $\mathrm{SFR} = 100~\mathrm{M}_{\odot}\mathrm{yr}^{-1}$ and normalized to unity. According to Eq.~\ref{mtotsfrdt}, both cases give the same total mass in the end, namely $M_{\mathrm{tot}} = 10 \cdot 10~\mathrm{M}_{\odot}\mathrm{yr}^{-1} \cdot \delta t = 1 \cdot 100~\mathrm{M}_{\odot}\mathrm{yr}^{-1} \cdot \delta t$. However, as can be seen in Fig.~\ref{fig_intro_plot}, cf.~panel (Ic) and (IIc), the distribution of the masses differ considerably since in the second case SCs of much higher masses can form due to the higher SFR.

As shown exemplarily above, any SFH can be converted to an $F(\mathrm{SFR})$. Conversely, the SFH cannot be derived from a known $F(\mathrm{SFR})$ because $F(\mathrm{SFR})$ does not reveal the chronological order of the involved SFRs. 

\section{Determining the star formation duration $\boldsymbol{\delta} \boldsymbol{t}$ using the SFR-$\boldsymbol{M_{\mathrm{ecl,max}}}$ relation} \label{sect_dt}

The SFR-$M_{\mathrm{ecl,max}}$ relation originates from a relation between the global SFR of a galaxy and the brightest SC in the $V$ band and was found by \citet{larsen02}. Later, \citet{adamo11} investigated the same properties of massive young SCs in blue compact galaxies and found them to lie slightly above the upper end of the Larsen relation. Moreover, \citet{randria13} observed a similar relation in the near-infrared for the brightest super star clusters (SSCs) in luminous infrared galaxies.

\begin{figure}[t]
\includegraphics[angle=-90, width=0.488\textwidth]{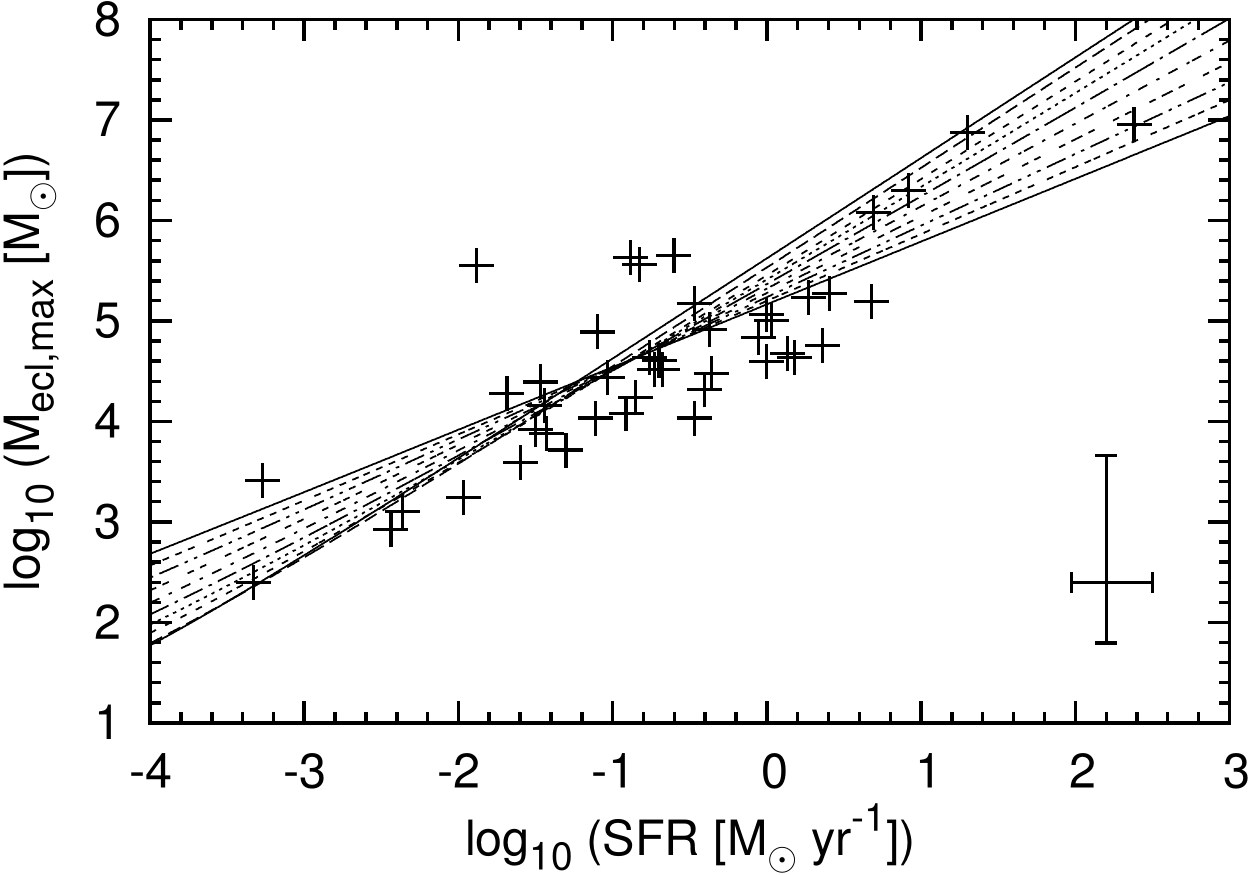}
\caption[SFR-$M_{\mathrm{ecl,max}}$ relation]{$M_{\mathrm{ecl,max}}$ vs. galaxy-wide SFR (replotted from \citet{weidner04}, including new data points). The curves are least-squares fits according to the SFR-$M_{\mathrm{ecl,max}}$ relation (Eq.~\ref{sfr_M_eclmax}) with the fitting parameter $\delta t$ for all $\beta$ between 1.5 (steepest curve) and 2.6 (shallowest curve) in steps of 0.1. The curves for $\beta = 1.6$, and 1.8 are omitted. A typical error estimate is indicated in the bottom right corner.}
\label{fig_sfr_Mmax5}
\end{figure}

In an analysis based on \citet{larsen02}, \citet{weidner04} showed the brightest SC to be the most massive young SC in most cases, even though the $M$/$L$-ratio of a stellar population depends highly on its age. They converted the SFR-brightest SC relation to the SFR-$M_{\mathrm{ecl,max}}$ relation and analyzed it for dwarf and spiral galaxies. Their data (young, most massive SCs ($M_{\mathrm{ecl,max}}$) vs.~current, galaxy-wide SFR, taken from \citealt{larsen02}; \citealt{weidner04}; \citealt{larsen09}; provided by C. Weidner, private communication) are replotted in Fig.~\ref{fig_sfr_Mmax5} and confirm a correlation between the SFR and $M_{\mathrm{ecl,max}}$. A typical error estimate is plotted in the bottom right corner. On the x-axis, the calibration of the SFR as a function of the infrared flux is the main contributor to the error, while on the y-axis the uncertainties mostly originate from converting the luminosity of a SC into a mass with an assumed $M$/$L$-ratio that strongly depends on the age of the SC (C. Weidner, private communication, see also \citealt{weidner04}). 

Apparently, there is some spread in the data, particularly above the relation. These data points mostly belong to dwarf and irregular starburst galaxies \citep{billett02,larsen02}. Several explanations have been suggested for this offset: \citet{weidner04} argued that an intense star formation activity in a dwarf galaxy may be halted once a very massive SC has formed (the "quenching" hypothesis). The feedback of this SC may heat the surrounding dense gas and prevent further star formation. This scenario was supported by \citet{bastian08} based on a study of NGC~1569, a dwarf-irregular (post) starburst galaxy. According to the quenching hypothesis, the outlying data points may be located at incorrect positions in the diagram: The SFR might have dropped significantly after the formation of these SCs, for which reason the data points would have to be horizontally shifted to the right since they formed at higher SFRs and may lie in the area of the other measurements.

On the other hand, for the Milky Way (MW) galaxy and the Large Magellanic Cloud (LMC), \citet{fukui99} suggested that the stronger gravitational field in the MW compared to the LMC leads to a stronger fragmentation of molecular clouds, for which reason the MW is able to form solely open SCs. In contrast, the weaker gravitational field of the LMC allows the formation of more populous SCs. More generally, \citet{billett02} proposed that dwarf galaxies are able to form massive SSCs due to the absence of shear. \citet{weidner10} investigated how shear forces act on GMCs in dwarf and spiral galaxies. They found that the presence of shear prevents GMCs to collapse into dense SSCs in spirals, while in dwarfs the lack of rotational support allows the formation of SSCs. Thus, dwarf galaxies would be able to form more massive SCs than spirals at the same SFR and therefore lie above the mentioned SFR-$M_{\mathrm{ecl,max}}$ relation. This would be equivalent to $\xi_{\mathrm{ECMF}} (M)$ deviating from the canonical form (Eq.~\ref{ecmf}) for some dwarf galaxies.

To begin, all data points are included, but we examine below how the analysis is influenced when the four data points lying above the relation in Fig.~\ref{fig_sfr_Mmax5} with $M_{\mathrm{ecl,max}}$ between $10^5~\mathrm{M}_{\odot}$ and $10^6~\mathrm{M}_{\odot}$ and $\log_{10} (\mathrm{SFR}) < 0$ are excluded. These data points belong to measurements in NGC~1705, NGC~1569, the Small Magellanic Cloud, and the LMC, viewed from left to right. However, assuming that these data points are placed at the correct positions, this would require an SC formation timescale of at least 10~Myr according to a simple estimate using Eq.~\ref{mtotsfrdt}.

The length of one SC formation epoch, $\delta t$, was determined by fitting the SFR-$M_{\mathrm{ecl,max}}$ relation (Eq.~\ref{sfr_M_eclmax}) to all data points using the least-squares method. Since $\delta t$ might vary with $\beta,$ a fit for each $\beta$ was performed separately in the range from 1.5 to 2.6 in steps of 0.1. Figure~\ref{fig_sfr_Mmax5} shows the fitted curves through the data points. The steepest curve corresponds to $\beta = 1.5$, the shallowest curve to $\beta = 2.6$. For purposes of clarity, the curves belonging to $\beta = 1.6$, and 1.8 are omitted. 

\begin{table}[tb]
\caption[Duration of one SC formation epoch, $\delta t$, as a function of $\beta$]{Duration of one SC formation epoch, $\delta t$, and the reduced $\chi^2_{\mathrm{red}}$ as determined from the least-squares fits in Fig.~\ref{fig_sfr_Mmax5} for each $\beta$ (column~1). In columns~2 and 3 all data points are used while in column~4 and 5 the four data points lying above the relation in Fig.~\ref{fig_sfr_Mmax5} (with $10^5~\mathrm{M}_{\odot} \lesssim M_{\mathrm{ecl,max}} \lesssim 10^6~\mathrm{M}_{\odot}$) are excluded. All values are visualized in Fig~\ref{fig_beta_dt_chi}.}
\label{tab_deltat}
\centering
\begin{tabular}{c|cc|cc}
\hline \hline
$\beta$ 			&       $\delta t$      	& $\chi^2_{\mathrm{red}}$	&       $\delta t$     	& $\chi^2_{\mathrm{red}}$	\\
				&       [Myr]   		&        	&       		[Myr]   	&             \\
\hline
1.5  				&       0.42            	&       0.409	&       0.31            		&       0.230           \\
1.6     			&       0.55    		&       0.409	&       0.41            		&       0.229           \\
\cellcolor[gray]{0.9}1.7        &       0.77            	&       0.409	&       0.56            		&       0.226           \\
\cellcolor[gray]{0.9}1.8        & \cellcolor[gray]{0.9}1.11	&       0.410	&       0.81            		&       0.223           \\
\cellcolor[gray]{0.9}1.9        & \cellcolor[gray]{0.9}1.70	&       0.414	& \cellcolor[gray]{0.9} 1.24            &       0.219           \\
\cellcolor[gray]{0.8}2.0        & \cellcolor[gray]{0.8}2.80	&       0.424	& \cellcolor[gray]{0.8} 2.01            &       0.216           \\
\cellcolor[gray]{0.8}2.1        & \cellcolor[gray]{0.8}4.94	&       0.443	& \cellcolor[gray]{0.8} 3.51            &       0.218           \\
\cellcolor[gray]{0.8}2.2        & \cellcolor[gray]{0.8}9.31	&       0.475	& \cellcolor[gray]{0.8} 6.51            &       0.230           \\
\cellcolor[gray]{0.8}2.3        & \cellcolor[gray]{0.9}18.57	&       0.524	& \cellcolor[gray]{0.8} 12.75           &       0.255           \\
\cellcolor[gray]{0.9}2.4        &       38.77           	&       0.591	& \cellcolor[gray]{0.9} 26.09           &       0.296           \\
2.5     			&       83.77           	&       0.677	&       55.19           		&       0.355           \\
2.6     			&       185.82          	&	0.782	&       119.79          		&       0.432           \\
\hline
\end{tabular}
\end{table}

\begin{figure}[tb]
\includegraphics[angle=-90, width=0.488\textwidth]{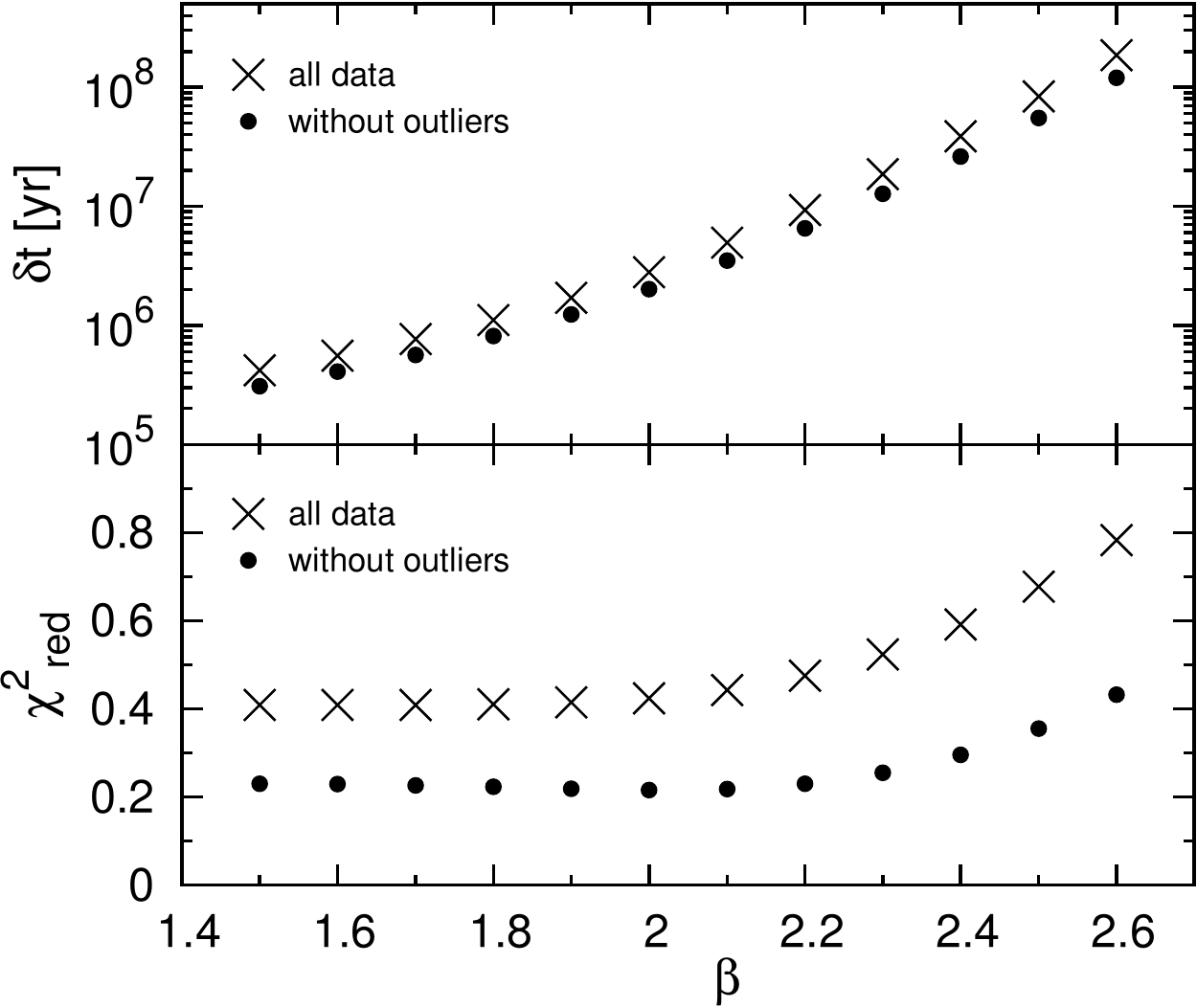}
\caption[Duration of one SC formation epoch, $\delta t$, and the reduced $\chi^2_{\mathrm{red}}$ as a function of $\beta$]{Duration of one SC formation epoch, $\delta t$, (upper panel) and the reduced $\chi^2_{\mathrm{red}}$ (lower panel) as determined from the fitting in Fig.~\ref{fig_sfr_Mmax5} with (crosses) and without the four outliers (filled circles) for each $\beta$. The values can be found in Table~\ref{tab_deltat}.}
\label{fig_beta_dt_chi}
\end{figure}

Clearly, $\delta t$ increases with increasing $\beta$, ranging from 0.4~Myr to 186~Myr (Table~\ref{tab_deltat}, column~2), as visualized by the crosses in the upper panel of Fig.~\ref{fig_beta_dt_chi}. A physical explanation might be that it takes longer to populate an ECMF with a large $\beta$ than an ECMF with a small $\beta$ due to the larger number of SCs for a given $M_{\mathrm{max}}$ (cf.~Eq.~\ref{N_ecmf}). In addition, the reduced $\chi_{\mathrm{red}}^2$ values are extracted from the fit analysis (Table~\ref{tab_deltat}, column~3) and indicated by crosses in the lower panel of Fig.~\ref{fig_beta_dt_chi}. Regarding $\chi_{\mathrm{red}}^2$, the most probable combinations of $\beta$ and $\delta t$ are those for $\beta \lesssim 2$. Toward larger $\beta$, $\chi_{\mathrm{red}}^2$ increases but does not exceed unity.

For comparison, the same fitting analysis is applied to the SFR-$M_{\mathrm{ecl,max}}$ data set (Fig.~\ref{fig_sfr_Mmax5}) excluding the above mentioned four data points lying above the relation. The resulting length of one SC formation epoch, $\delta t$, can be viewed in column~4 in Table~\ref{tab_deltat} and is represented by filled circles in the upper panel of Fig.~\ref{fig_beta_dt_chi}. It emerges that the values for $\delta t$ are somewhat lower than the previous fitting results (cf.~columns~2 and 4 in Table~\ref{tab_deltat}, Fig.~\ref{fig_beta_dt_chi}). This is anticipated since $\delta t$ can be found in the denominator of the SFR-$M_{\mathrm{ecl,max}}$ relation (Eq.~\ref{sfr_M_eclmax}). Consequently, an increasing $\delta t$ induces the fitting function to shift downward in Fig.~\ref{fig_sfr_Mmax5}. However, fitting Eq.~\ref{sfr_M_eclmax} to the data without the outliers already places this function slightly below the previous fits since the outliers lie above the actual relation. Thus, $\delta t$ is smaller than before. Moreover, the removal of the outliers leads to a smaller spread in the remaining data set and therefore to lower reduced $\chi_{\mathrm{red}}^2$ values than in the previous fitting. This becomes clear from the lower panel of Fig.~\ref{fig_beta_dt_chi}, where the obtained $\chi_{\mathrm{red}}^2$ values are marked with filled circles, and from comparing columns~3 and 5 in Table~\ref{tab_deltat}.

The theoretical star formation timescale of about 10~Myr follows from the calculation of the Jeans time in molecular clouds \citep[e.g.,][]{egusa04}. This value has been widely adopted \citep[e.g.,][]{billett02, weidner04, maschberger07}. In simulations performed by \citet{bonnell06}, star formation occurs within 2~Myr, while the surrounding cloud disperses on a timescale of 10~Myr.

There have been various attempts to estimate the timescale of SC formation observationally: From analyzing GMCs in the LMC, \citet{fukui99} estimated SC formation to proceed in a few Myr and a typical lifetime of a molecular cloud of about 6~Myr. Likewise in the LMC, \citet{yamaguchi01} found that SCs are actively formed over roughly 4~Myr and the host molecular clouds completely dissipates in about 10~Myr after the onset of SC formation. Another approach to observationally estimate the SC formation timescale is to measure the offset between H$\alpha$, emitted through recombination of hydrogen ionized by newly born massive stars, and CO, a tracer of molecular gas which is observed in star-forming spiral arms. Star formation times derived in this way by \citet{egusa04, egusa09} range from 4 to 28~Myr, whereas half of the measurements lie between 11 and 14~Myr. Similarly, \citet{tamburro08} compared images of spiral galaxies in H{\footnotesize I} from cold gas and 24~$\mu$m from warm dust heated by UV and find shorter timescales between 1 and 4~Myr.

By combining the results from theory, simulations, and observations, we find agreement that the formation of an SC population through the formation and the dispersal of their birth molecular clouds occurs galaxy-wide on a timescale between at least a few Myr and at most a few 10~Myr. Our fitting results for $\delta t$, the duration of one SC formation epoch, which match these estimates, are highlighted in light gray, while the most probable values are shaded slightly darker in Table~\ref{tab_deltat}. Additionally, observed values for $\beta$, the index of the ECMF, are presented in light gray, while the values found most frequently are highlighted somewhat darker. As one can see immediately, the colored entries in the two columns overlap over almost the full range. This demonstrates the reliability of the analytically derived SFR-$M_{\mathrm{ecl,max}}$ relation (Eq.~\ref{sfr_M_eclmax}) since it naturally connects -- without any adjustment -- the empirical estimates of the two independent quantities $\beta$ and $\delta t$ in combination with the SFR vs.~$M_{\mathrm{ecl,max}}$ data from \citet{weidner04}. Note that this finding is virtually independent of whether outliers are excluded or not (cf.~Table~\ref{tab_deltat}).

Since there has not been a definitive statement about the four outliers and the outliers do not change the results much, none of them is excluded. Thus, for all further calculations we use the values for $\delta t$ as given in column~2 in Table~\ref{tab_deltat}. However, it is not expected that the further analysis will depend much on whether the outliers are excluded or not since in both cases the values for $\delta t$ are similar and increase with $\beta$ in a similar way (cf.~columns~2 and 4 in Table~\ref{tab_deltat}, Fig.~\ref{fig_beta_dt_chi}). Moreover, we assumed that the SFR-$M_{\mathrm{ecl,max}}$ relation, extrapolated to higher values, holds true. 

\section{Considered star formation histories (SFHs)} \label{sect_SFH}

An SFH reveals how the SFR evolves with time, that means $\mathrm{SFH} \equiv \mathrm{SFR} (t)$. Various observations suggest that on cosmological scales, the SFR was higher in the past or for higher redshifts, $z$, than it is today \citep[e.g.,][see also references therein]{schiminovich05, lefloch05}, meaning that $\mathrm{SFR} (t)$ must decrease with time. For instance, this was observed for the MW by \citet{kroupa02}, for the individual galaxies NGC~584, NGC~3377, and NGC~3610 by \citet{georgiev12} and for a sample of massive galaxies by \citet{daddi07}.

The simplest and most often considered SFH is a purely exponential SFH. It is characterized by an SFR that has its maximum at the beginning and exponentially decreases thereafter. A "delayed-exponential" SFH was suggested by \citet{sandage86}. Starting with $\mathrm{SFR} = 0$, the SFR increases to a maximum and exponentially decreases thereafter. Field galaxies show a similar development in the SFR between a redshift $z \approx 4$ and today, $z=0$, with a peak at $z \approx 1.5$ \citep{madau98}. The two described exponential SFHs were used by \citet{gavazzi02} and are considered here as well. 

Moreover, we also examined four power-law SFHs, which were chosen arbitrarily. They must have a similar evolution with $t$ as an exponential SFH, namely starting at a relatively high SFR value and monotonically decreasing from there. Also, they were required to be easy to handle. Thus, the power-law SFHs have to be of the form $\mathrm{SFR} (t) \propto t^{- \eta}$. To investigate different indices, $\eta = 0.5,$ 1, 2, and 3 were considered. Interestingly, \citet{lilly96} found that the comoving luminosity density of the Universe can be reasonably well described by a power-law SFH with $\eta = 2.5$.

The examined SFHs and their individual normalization constants, $c$, are

\paragraph{Exponential SFH,} 

  \begin{equation} \label{exp_sfh}
      \mathrm{SFR} (t) = \frac{c_{\mathrm{exp}}}{\tau_{\mathrm{exp}}} \exp \left( - \frac{t}{\tau_{\mathrm{exp}}} \right) ~ ; \quad c_{\mathrm{exp}} = M_{\mathrm{tot}} ,
  \end{equation}

\noindent $\tau_{\mathrm{exp}}$ parameterizes how fast the SFR decreases with $t$. 

\paragraph{Delayed-exponential SFH,} 

  \begin{equation}
      \mathrm{SFR} (t) = c_{\mathrm{del}} \frac{t}{\tau^2_{\mathrm{del}}} \exp \left( - \frac{t^2}{2 \tau^2_{\mathrm{del}}} \right) ~ ; \quad c_{\mathrm{del}} = M_{\mathrm{tot}} ,
  \end{equation}

\noindent $\tau_{\mathrm{del}}$ parameterizes at which time the SFR reaches maximum.

\paragraph{Power-law SFH with $\eta = 0.5$,}

  \begin{equation}
      \mathrm{SFR} (t) = c_{0.5} ~ t^{-0.5} ~ ; \quad c_{0.5} = \frac{M_{\mathrm{tot}}}{2 \sqrt{t_{\mathrm{Hubble}}}} ,
  \end{equation}

\paragraph{Power-law SFH with $\eta = 1$,}

  \begin{equation}
      \mathrm{SFR} (t) = c_{1} ~ t^{-1} ~ ; \quad c_{1} = \frac{M_{\mathrm{tot}}}{\ln t_{\mathrm{Hubble}} - \ln \delta t} ,
  \end{equation}

\paragraph{Power-law SFH with $\eta = 2$,}

  \begin{equation}
      \mathrm{SFR} (t) = c_{2} ~ t^{-2} ~ ; \quad c_{2} = M_{\mathrm{tot}} \cdot \delta t ,
  \end{equation}

\paragraph{Power-law SFH with $\eta = 3$,}

  \begin{equation} \label{power3}
      \mathrm{SFR} (t) = c_{3} ~ t^{-3} ~ ; \quad c_{3} = 2 ~ M_{\mathrm{tot}} \cdot \delta t^2 .
  \end{equation}

All the above SFHs are normalized such that the integration of the SFH ($\equiv \mathrm{SFR} (t)$) yields the total stellar mass, $M_{\mathrm{tot}}$, of all SCs that ever formed:

  \begin{equation} \label{norm_crit}
      M_{\mathrm{tot}} = \int_{0}^{\infty} \mathrm{SFR} (t) ~ \mathrm{d} t .
  \end{equation}

\noindent This normalization criterion cannot be applied to some of the SFHs because the lower, $t = 0$, and/or the upper limit, $t = \infty$, would lead to a diverging integral. If this was the case, the lower and/or the upper limit were replaced by $t = \delta t$ and/or $t = t_{\mathrm{Hubble}}$, respectively, in Eq.~\ref{norm_crit}. For the exponential and the delayed-exponential SFH, the parameters $\tau_{\mathrm{exp}}$ and $\tau_{\mathrm{del}}$, respectively, were chosen to be 1~Gyr, which means that the predominant star formation activity concentrates on the first few Gyr,

   \begin{equation} \label{tau}
       \tau_{\mathrm{exp}} = \tau_{\mathrm{del}} = 1~\rm{Gyr} .
  \end{equation}

We assumed that over a Hubble time, SCs are formed with a total stellar mass of 

   \begin{equation} \label{norm_SFH}
        M_{\mathrm{tot}} = 10^{10}~\mathrm{M}_{\odot} .
  \end{equation}

\noindent This amount is compatible with the total mass stored in SCs systems in today's galaxy clusters. For instance, a simple estimate from the mass distribution of GCs and ultra-compact dwarf galaxies (UCDs) in the Fornax galaxy cluster by \citet{hilker09} leads to approximately this mass. However, without loss of generality, one can assume any value for $M_{\mathrm{tot}}$. The assumed mass estimate does not necessarily imply that the assembly of all these SCs will host the same amount of mass today. Since evolutionary effects are not taken into account here, mass loss through stellar evolution does not occur. In real SCs, massive stars lose gas due to stellar winds or supernova explosions, which becomes available again for later generations of SCs. Thus, even if all SCs ever formed had a mass of $10^{10}~\mathrm{M}_{\odot}$ at their birth in total, the agglomeration of them will be significantly less massive today since SCs lose about 90~\% of their birth stellar mass within several Gyr, while those with $M_{\mathrm{ecl}} \lesssim 10^{4}~\mathrm{M}_{\odot}$ dissolve completely.

\section{Determining $\boldsymbol{F(\mathrm{SFR})}$ from the considered SFHs} \label{sect_F}

According to the following procedure, the distribution functions of SFRs, $F(\mathrm{SFR})$, were calculated from the considered SFHs:

\begin{enumerate}
\item Determination how many SC formation epochs occurred: Each SFH (Sect.~\ref{sect_SFH}) is interpreted to be a sequence of single SC formation epochs and is divided into $N_{\mathrm{SCFE}} = t_{\mathrm{Hubble}} / \delta t$ epochs of length $\delta t$ as sketched in the left panel of Fig.~\ref{fig_sketch}. \label{step1}
\item Calculation of the average SFR for each SC formation epoch: The SFR is evaluated at the beginning and at the end of each formation epoch. Moreover, the integral of the $n$-th SC formation epoch

  \begin{equation}
      \int_{n \cdot \delta t}^{(n+1) \cdot \delta t} \mathrm{SFR} (t) ~ \mathrm{d} t = \underbrace{ \int_{n \cdot \delta t}^{(n+\kappa) \cdot \delta t} \mathrm{SFR} (t) ~ \mathrm{d} t }_{I_1} + \underbrace{ \int_{(n+\kappa) \cdot \delta t}^{(n+1) \cdot \delta t} \mathrm{SFR} (t) ~ \mathrm{d} t }_{I_2} ,
  \end{equation} 

\noindent is divided into two integrals and the limit $(n+\kappa) \cdot \delta t$ is chosen such that the two integrals are equal, that is, $I_1 = I_2$. This ensures that Eq.~\ref{mtotsfrdt} is fulfilled for each SC formation epoch. Then the average SFR of the $n$-th SC formation epoch is evaluated for each SFH (Eqs.~\ref{exp_sfh} - \ref{power3}) from $\mathrm{SFR}_{\mathrm{av},n} = \mathrm{SFR} ((n+\kappa) \cdot \delta t)$. $\kappa$ is individually determined for each SC formation epoch. \label{step2}
\item Creating a histogram $N_{\mathrm{SCFE}} (\mathrm{SFR})$: Such a histogram reveals how often every $\mathrm{SFR}$ occurred.

\begin{figure}[t]
\includegraphics[angle=-90, width=0.488\textwidth]{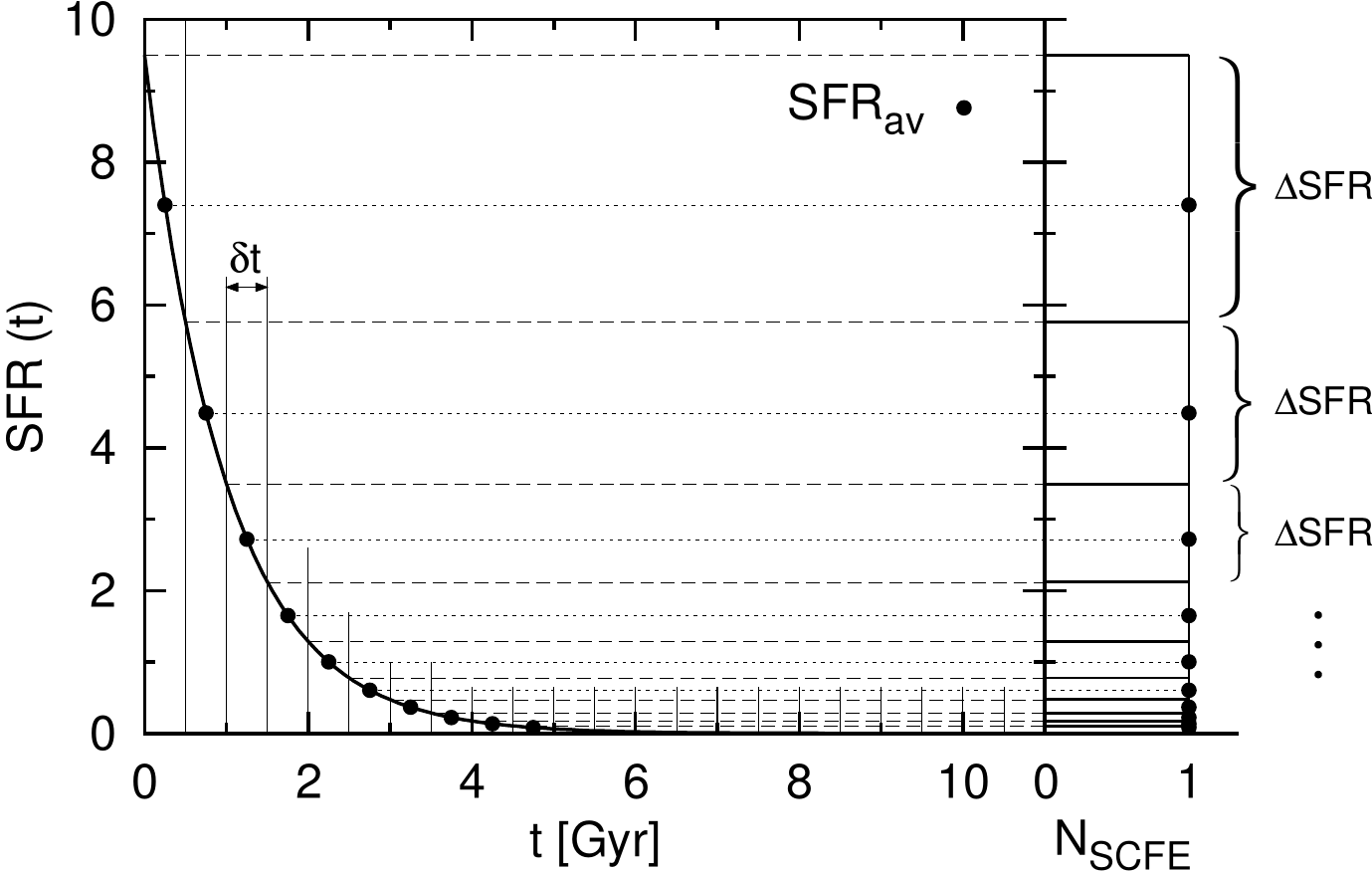}
\caption[Sketch: How $F(\mathrm{SFR})$ is determined from a SFH]{Sketch of how $F(\mathrm{SFR})$ is determined from a SFH: $\mathrm{SFR} (t)$ (left panel, here an exponentially declining SFH, Eq.~\ref{exp_sfh}) is divided into single SC formation epochs of length $\delta t$. An average $\mathrm{SFR}_{\mathrm{av}}$ is determined for each of the $N_{\mathrm{SCFE}}$ formation epochs. From this, a histogram (right panel) is created in which every $\mathrm{SFR}_{\mathrm{av}}$ has its own bin of height unity. $F(\mathrm{SFR})$ for each $\mathrm{SFR}_{\mathrm{av}}$ is determined by dividing unity by the width of the corresponding bin, $\Delta \mathrm{SFR}$.}
\label{fig_sketch}
\end{figure}

If an SFH is a monotonically decreasing function, each $\mathrm{SFR}_{\mathrm{av},n}$ occurs once. In this case, for each SC formation epoch, one bin of height unity is created at $\mathrm{SFR}_{\mathrm{av},n}$. The width of the bin is defined by the SFRs at the beginning and at the end of that SC formation epoch. This is illustrated in the right panel of Fig.~\ref{fig_sketch}: If the x- and y-axis are exchanged, the histogram $N_{\mathrm{SCFE}} (\mathrm{SFR})$ is depicted in the usual manner. 

If an SFH is not a decreasing function, $\mathrm{SFR}_{\mathrm{av},n}$ are sorted according to their value. To create the same number of bins as there are $\mathrm{SFR}_{\mathrm{av},n}$ values, the upper and lower limit of the $n$-th bin is computed from taking the arithmetic average between the $\mathrm{SFR}_{\mathrm{av},n}$ and the previous and the next value: $(\mathrm{SFR}_{\mathrm{av},n} + \mathrm{SFR}_{\mathrm{av},n \pm 1}) / 2$.

\begin{figure*}[t]
\includegraphics[angle=-90, width=\textwidth]{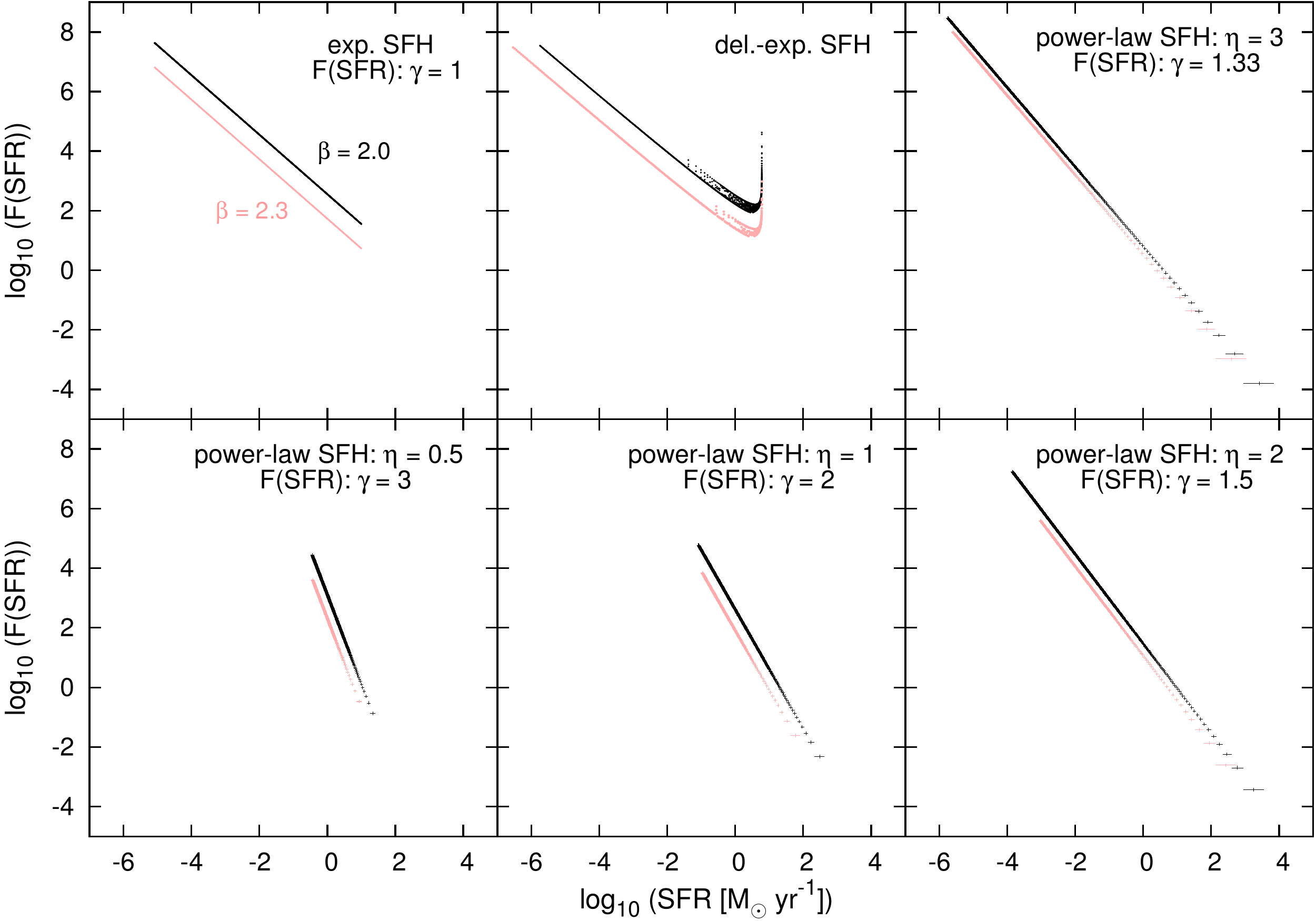}
\caption[$F(\mathrm{SFR})$ of the considered SFHs for $\beta = 2.0$ and $\beta = 2.3$]{$F(\mathrm{SFR})$, the distribution function of SFRs, of the two exponential and four power-law SFHs (Eqs.~\ref{exp_sfh} - \ref{power3} with Eqs.~\ref{tau} - \ref{norm_SFH}) for $\beta = 2.0$ (black) and $\beta = 2.3$ (red). Each data point corresponds to one SC formation epoch. Error bars are indicated if they are large enough to display. Except for the delayed-exponential SFH, all $F(\mathrm{SFR})$ can be exactly fitted by a power law according to Eq.~\ref{fit}. The corresponding index $\gamma$ and the considered SFH are given in the corner of each panel. The panels are arranged such that $\gamma$ increases from the upper left to the lower left by viewing in clockwise direction. (A color version of this figure is available in the online journal.)}
\label{fig_F_20+23}
\end{figure*}

Pursuant to this notion, $N_{\mathrm{SCFE}}$ bins of height unity are obtained without any gaps in between. This enables determining $F(\mathrm{SFR})$ with the highest possible precision because the number of bins in the histogram is equal to the number of SC formation epochs. \label{step3}
\item Calculation of $F(\mathrm{SFR})$: For each $\mathrm{SFR}_{\mathrm{av},n}$, unity is divided by the width of the $n$-th bin, $\Delta \mathrm{SFR}$, according to Eq.~\ref{F}. Since $M_{\mathrm{min}} = 5~\mathrm{M}_{\odot}$, only those SFRs are considered that lead to the formation of SCs more massive than $5~\mathrm{M}_{\odot}$. \label{step4}
\end{enumerate}

For the six considered SFHs from Sect.~\ref{sect_SFH} (Eqs.~\ref{exp_sfh} - \ref{power3} with Eqs.~\ref{tau} - \ref{norm_SFH}), $F(\mathrm{SFR})$ is determined as described above. This is done for all $\beta$ of Table~\ref{tab_deltat} because the results may vary with $\delta t$. Interestingly, the essential properties of $F(\mathrm{SFR})$ do not change much with $\beta$, and thus $F(\mathrm{SFR})$ of the six SFHs is shown only for $\beta = 2.0$ in black and for $\beta = 2.3$ in red in Fig.~\ref{fig_F_20+23}. In all cases, $F(\mathrm{SFR})$ is composed of many individual data points, each of them corresponding to one SC formation epoch. Error bars are indicated if they are larger than the size of the used plotting symbol.

\begin{figure}[htb]
\includegraphics[angle=-90, width=0.488\textwidth]{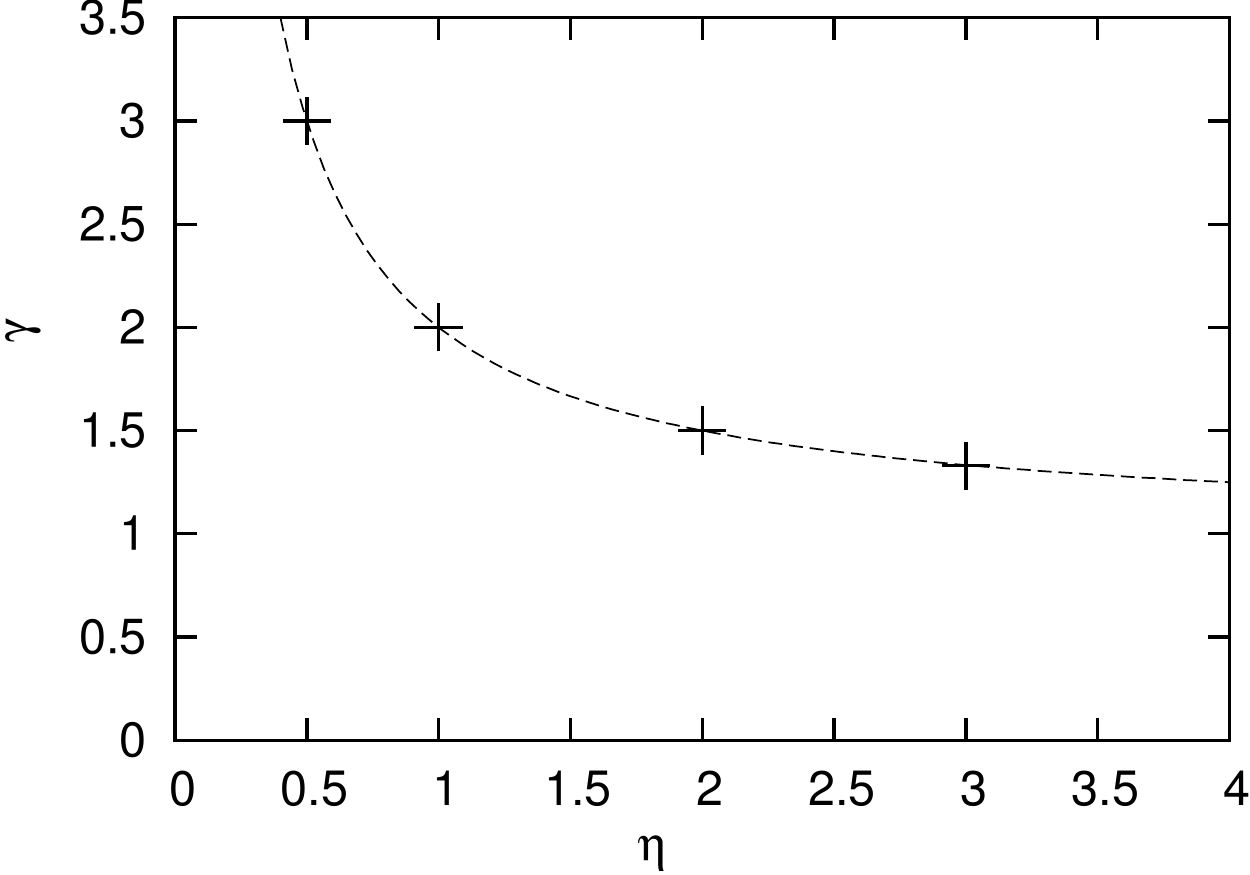}
\caption[$F(\mathrm{SFR})$ index $\gamma$ as a function of the SFH index $\eta$]{$F(\mathrm{SFR})$ index $\gamma$ as a function of the index $\eta$ of the power-law SFHs. The indicated curve, $\gamma = 1 + 1/\eta$, exactly matches the data points.}
\label{fig_slopes}
\end{figure}

It becomes immediately apparent that all $F(\mathrm{SFR})$ -- except for a delayed-exponential SFH, which is discussed below -- are basically straight lines in the double logarithmic representation. This means that the quantity SFR itself is distributed according to a power law independent of whether the SFH ($\equiv \mathrm{SFR} (t)$) is an exponential or a power law. Power laws according to 

   \begin{align} \label{fit}
\begin{split}
     \log_{10} (F(\mathrm{SFR})) & = - \gamma \log_{10} (\mathrm{SFR}) + g \\
 \Leftrightarrow \quad \qquad F(\mathrm{SFR}) & = 10^g \cdot \mathrm{SFR}^{- \gamma} ,
\end{split}
   \end{align}

\noindent were fitted to these $F(\mathrm{SFR}),$ from which the fitting parameters $\gamma$ and $g$ were determined. Figure~\ref{fig_F_20+23} visualizes a slight vertical shift between $F(\mathrm{SFR})$ obtained for $\beta = 2.0$ (black) and $\beta = 2.3$ (red), meaning that the y-intercept $g$ changes with $\beta,$ but the slope $- \gamma$ does not. Even if the normalization constants are varied by choosing a higher or lower total mass, $M_{\mathrm{tot}}$, (Eq.~\ref{norm_crit}) or another value for the parameter $\tau_{\mathrm{exp}}$ (Eq.~\ref{tau}) in the exponential SFH (Eq.~\ref{exp_sfh}) -- the values of $\gamma$ are unaffected. In Fig.~\ref{fig_F_20+23} the six panels containing the resulting $F(\mathrm{SFR})$ of the six SFHs are arranged such that the index $\gamma$ of $F(\mathrm{SFR})$ -- as indicated in the upper right corner of each panel -- increases if the panels are viewed in clockwise direction, starting at the upper left panel and ending at the lower left panel. In Table~\ref{tab_alpha_vs_gamma} the five SFHs whose $F(\mathrm{SFR})$ follow a power law and their respective index $\gamma$ are collated. In addition, the four power-law SFH indices $\eta$ are plotted against their respective $F(\mathrm{SFR})$ indices $\gamma$ in Fig.~\ref{fig_slopes}. Interestingly, these data points can exactly be fitted by the function $\gamma = 1 + 1/\eta$. All this indicates a fundamental relation between the type of SFH and the index of the its $F(\mathrm{SFR})$.

\begin{table}[htb]
\caption[The SFH vs. the corresponding $F(\mathrm{SFR})$ index $\gamma$]{SFH and the corresponding index $\gamma$ of $F(\mathrm{SFR})$}
\label{tab_alpha_vs_gamma}
\centering
\begin{tabular}{ll}
\hline \hline
SFH                             &       $\gamma$        \\
\hline
exponential                     &       1.0             \\
power law, $\eta = 3$           &       1.33            \\
power law, $\eta = 2$           &       1.5             \\
power law, $\eta = 1$           &       2.0             \\
power law, $\eta = 0.5$         &       3.0             \\
\hline
\end{tabular}
\end{table}

Figure~\ref{fig_F_20+23} shows that the $F(\mathrm{SFR})$ of the delayed-exponential SFH again increases toward the high-SFR end. This feature is caused by the shape of the delayed-exponential SFH: The SFR increases first and then decreases, so that relatively high SFRs appear considerably more often than they do in a purely exponential SFH, which is a decreasing function at any time. As a result of the large number of SC formation epochs at high SFRs, $F(\mathrm{SFR})$ at high SFRs is enhanced. After the delayed-exponential SFH has reached its maximum, it decreases exponentially in the same way as a purely exponential SFH. Within the regime of low SFRs, $F(\mathrm{SFR})$ can thus be described by a power law as well that has almost exactly the same index $\gamma$ as $F(\mathrm{SFR})$ of a purely exponential SFH. Therefore, the panel containing the $F(\mathrm{SFR})$ of the delayed-exponential SFH is placed to the right of the one with the purely exponential SFH.

Indeed, the relation between $\eta$ and $\gamma$ is fundamental and can be derived analytically in the following way: $F(\mathrm{SFR})$ describes how the number of SC formation epochs, $N_{\mathrm{SCFE}}$, change with SFR. Since one is confronted with discrete numbers here, Eq.~\ref{F} needs to be discretized,

   \begin{equation} \label{F_discrete}
      F(\mathrm{SFR}) = \frac{\Delta N_{\mathrm{SCFE}} (\mathrm{SFR}) }{\Delta \mathrm{SFR}} = \frac{1}{ | \mathrm{SFR}^{\prime} (t) | \cdot \delta t } .
   \end{equation}

\noindent The last equality becomes clear using the sketch in Fig.~\ref{fig_sketch}: Each SFR occurs once during the SC formation episode, that is, $\Delta N_{\mathrm{SCFE}} = 1$. The change in SFR per unit time is by definition $\mathrm{SFR}^{\prime} (t) = \mathrm{d} \mathrm{SFR} (t) / \mathrm{d} t,$ so that $\Delta \mathrm{SFR} = | \mathrm{SFR}^{\prime} (t) | \cdot \delta t$. The absolute value of $\mathrm{SFR}^{\prime} (t)$ prevents $F(\mathrm{SFR})$ from being negative since the considered $\mathrm{SFR} (t)$ are decreasing functions. Moreover, $\mathrm{SFR} (t)$ needs to be a monotonic function. Otherwise, there would exist a maximum or a minimum turning point so that $\mathrm{SFR}^{\prime} (t)$ would become zero and Eq.~\ref{F_discrete} undefined. Therefore, this argumentation does not apply for the delayed-exponential SFH. As a result, in this case, $F(\mathrm{SFR})$ is not obtained analytically.

The derivatives of the exponential and of a general power-law SFH are 

   \begin{equation} \label{exp_prime}
      \mathrm{SFR}^{\prime} (t) = - \frac{c_{\mathrm{exp}}}{\tau_{\mathrm{exp}}^2} \exp \left( - \frac{t}{\tau_{\mathrm{exp}}} \right) = - \frac{1}{\tau_{\mathrm{exp}}} \cdot \mathrm{SFR} (t) ,
   \end{equation}

   \begin{equation} \label{pl_prime}
      \mathrm{SFR}^{\prime} (t) = - \eta c t^{-\eta - 1} = - {\tilde c} ( c t^{-\eta} )^{\frac{-\eta - 1}{-\eta}} = - {\tilde c} (\mathrm{SFR} (t))^{1 + \frac{1}{\eta}} .
   \end{equation}

\noindent In Eq.~\ref{pl_prime}, $\mathrm{SFR}^{\prime} (t)$ is rearranged such that $\mathrm{SFR} (t)$ appears again, while all constant factors are combined to $\tilde c$. Inserting Eq.~\ref{exp_prime} and Eq.~\ref{pl_prime}, respectively, in Eq.~\ref{F_discrete} gives

   \begin{equation} \label{F_exp}
   \begin{split}
      F(\mathrm{SFR}) & = \frac{\tau_{\mathrm{exp}}}{\mathrm{SFR} (t) \cdot \delta t} = \frac{{\tilde c}_{\mathrm{exp}}}{\mathrm{SFR} (t)} \\
      \Rightarrow \log_{10} ~ (F(\mathrm{SFR})) & = \log_{10} ({\tilde c}_{\mathrm{exp}}) - \log_{10} (\mathrm{SFR} (t)) ,
   \end{split}
   \end{equation}

   \begin{equation} \label{F_pl}
   \begin{split}
      F(\mathrm{SFR}) & = \frac{1}{{\tilde c} (\mathrm{SFR} (t))^{1 + \frac{1}{\eta}} \cdot \delta t} = \frac{{\tilde c}_{\mathrm{pow}}}{(\mathrm{SFR} (t))^{1 + \frac{1}{\eta}}} \\
      \Rightarrow \log_{10} ~ (F(\mathrm{SFR})) & = \log_{10} ({\tilde c}_{\mathrm{pow}}) - \left( 1 + \frac{1}{\eta} \right) \log_{10} (\mathrm{SFR} (t)) .
   \end{split}
   \end{equation}

\noindent One can immediately extract from Eq.~\ref{F_exp} that $F(\mathrm{SFR})$ derived from any exponential SFH always has a slope of $-\gamma = -1$. Deriving $F(\mathrm{SFR})$ from a power-law SFH (Eq.~\ref{F_pl}) leads to a slope of $-\gamma = -(1 + 1/\eta)$ in exact agreement with the above finding.

Because this derivation is not applicable to the delayed-exponential SFH, the corresponding $F(\mathrm{SFR})$ is not a straight line in a double logarithmic plot, meaning that it cannot be described by a power law, as already became clear from Fig.~\ref{fig_F_20+23}.

\begin{figure*}[t]
\includegraphics[angle=-90, width=\textwidth]{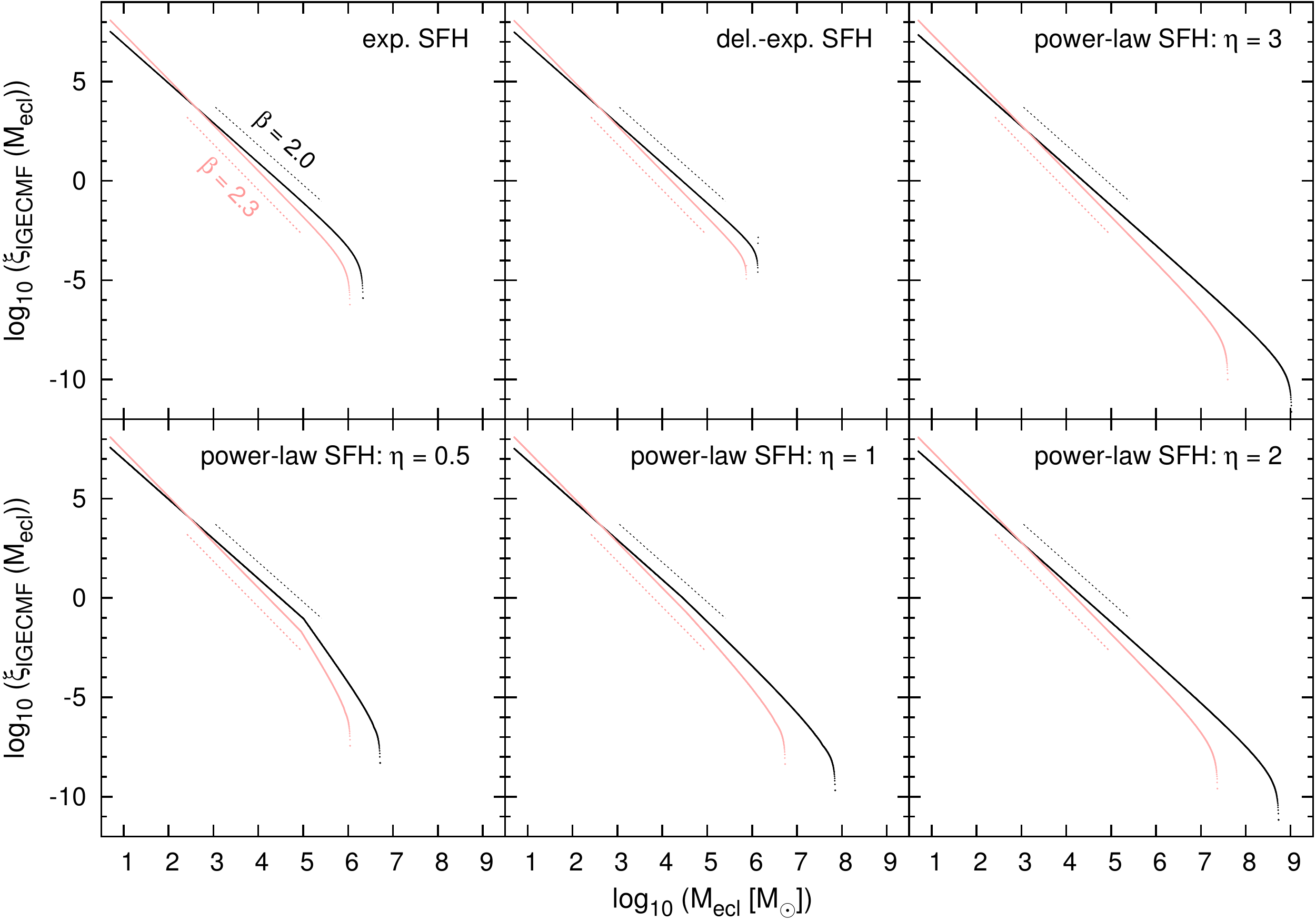}
\caption[The IGECMFs of the considered SFHs for $\beta = 2.0$ and $\beta = 2.3$]{IGECMFs for $\beta = 2.0$ (black) and $\beta = 2.3$ (red) showing the mass distribution function after a Hubble time of SC formation according to the SFH indicated in the corner of each panel. Strictly speaking, these IGECMFs represent how the birth stellar masses of SCs are distributed since evolutionary effects leading to a change in the SC mass are not taken into account. Shifted to higher or lower values, the dashed lines represent the underlying ECMFs in the same color for comparison. According to the underlying SFHs, the IGECMFs are arranged in the same manner as in Fig.~\ref{fig_F_20+23}. (A color version of this figure is available in the online journal.)}
\label{fig_igecmf_20+23}
\end{figure*}

\section{Deriving the IGECMFs from the obtained $\boldsymbol{F(\mathrm{SFR})}$} \label{sect_igecmf}

Now we can determine the IGECMFs from the derived $F(\mathrm{SFR})$ by evaluating Eq.~\ref{igecmf_mod}. For this, we transformed $F(\mathrm{SFR})$ to $F(M_{\mathrm{max}})$. This was done by converting the argument (SFR, x-axis in Fig.~\ref{fig_F_20+23}) to $M_{\mathrm{max}}$ according to Eq.~\ref{sfr}. The value of $F(\mathrm{SFR})$ is not affected by this rescaling. Then, we integrated Eq.~\ref{igecmf_mod} over all possible $M_{\mathrm{max}}$ from $M_{\mathrm{max}}^{\mathrm{low}} \equiv M_{\mathrm{min}} = 5~\mathrm{M}_{\odot}$ up to an individual $M_{\mathrm{max}}^{\mathrm{up}}$ that corresponds to the maximum SFR of the considered SFH. As with $F(\mathrm{SFR})$, the resulting IGECMFs for each SFH look very similar for different $\beta$. Thus, we show in Fig.~\ref{fig_igecmf_20+23} only the IGECMFs for $\beta = 2.0$ in black and for $\beta = 2.3$ in red. The resulting IGECMFs for one $\beta$ share several similarities that are independent of the underlying SFH:

\begin{itemize}
\item[\small$\bullet$] The starting position at the low-mass end is identical.
\item[\small$\bullet$] The shape of the lower mass part can be described by a power law with the slope of the underlying ECMF, represented by a dashed line in Fig.~\ref{fig_igecmf_20+23}.
\item[\small$\bullet$] The occurrence of a turn-down at the high-mass end: The IGECMFs become increasingly steeper toward the high-mass end, as compared to the parental ECMF. The IGECMFs are reminiscent of a Schechter function, which has a power-law behavior at the low-mass end and an exponential turn-down at the high-mass end. 
\end{itemize}

\noindent The turn-down shows several differences:
\begin{itemize}
\item[\small$\bullet$] The mass at which the bending down begins: This position is only especially distinct for the IGECMF resulting from the power-law SFH with $\eta = 0.5$. For all other IGECMFs, this point is not prominent because of a smooth transition.
\item[\small$\bullet$] The upper mass limit: It is directly related to the maximum SFR (Fig.~\ref{fig_F_20+23}) of the considered SFH through Eq.~\ref{sfr}.
\item[\small$\bullet$] The individual shape of the turn-down.
\end{itemize}

A comparison of the IGECMFs for $\beta = 2.0$ (black) and $\beta = 2.3$ (red) (Fig.~\ref{fig_igecmf_20+23}) shows that only the relative positions of the starting point at the low-mass end, the position where the turn-down appears, and the upper mass limit are shifted. The reason is that the SC masses are distributed differently depending on $\beta$: The SFR-$M_{\mathrm{ecl,max}}$ in Fig.~\ref{fig_sfr_Mmax5} shows that at a constant SFR a small $\beta$ corresponds to a large $M_{\mathrm{ecl,max}}$ and vice versa. This is directly reflected by the IGECMFs: For $\beta = 2.0$, the IGECMFs continue up to a higher upper mass limit than the IGECMFs with $\beta = 2.3$ (cf.~Fig.~\ref{fig_igecmf_20+23}). However, the same total mass, $M_{\mathrm{tot}}$, is stored in each IGECMF since all SFHs are normalized to the same $M_{\mathrm{tot}}$ (Eq.~\ref{norm_crit}) independent of $\beta$. To compensate for the higher upper mass limit, the low-mass tail of the IGECMFs with $\beta = 2.0$ is slightly shifted to lower values relative to the IGECMFs with $\beta = 2.3$, which is best visible at the low-mass end (cf.~Fig.~\ref{fig_igecmf_20+23}). Thus, assuming the same $M_{\mathrm{tot}}$, there will be fewer high-mass SCs and more low-mass SCs for an IGECMF with a large $\beta$ than for an IGECMF with a small $\beta$ -- exactly the same applies to the ECMF. The IGECMFs of all other $\beta$ exhibit the same features as listed above and are thus not shown here. 

The occurrence of an explicitly visible bend in the IGECMF -- as the one for the power-law SFH with $\eta = 0.5$ -- is the result of the underlying distribution of SFRs: The mass at which the bend appears corresponds to the upper mass limit, $M_{\mathrm{max}}$, of the lowest SFR (Eq.~\ref{sfr}). All SFRs contribute to the formation of SCs with masses below the mass at which the bend appears, which is why the IGECMF in this mass range has the same slope as the parental ECMF. Toward the high-mass end, a decreasing number of ECMFs contribute to the IGECMF since high SFRs become rarer. Thus, the IGECMF becomes increasingly steeper toward the high-mass end. For IGECMFs where a bend is not explicitly visible -- as in all other cases -- this explanation holds true as well.

The characteristic shape of the IGECMF -- a turn-down at the high-mass end and a power-law behavior in the mass range below -- is observed in the IGIMF \citep[e.g.,][]{kroupa_weidner03, weidner_kroupa05} for the same reason \citep{pflamm-altenburg07}, which means that all SFRs contribute at the low-mass range, while only the highest ones -- which are rare -- contribute to the high-mass range. This has a fundamental consequence: For an invariant ECMF, a mass distribution function accumulated over several formation epochs can only become steeper -- if high SFRs are rare -- or remain unchanged -- if high SFRs occur frequently -- toward the high-mass end, but it can never become shallower than the parental ECMF. Thus, the slope of the mass distribution function at the high-mass end gives clues on the slope of the parental ECMF: The underlying ECMF can be shallower or have the same slope as the observed mass function, but it cannot be steeper. Before the slope can be determined, however, it first has to be considered how stellar and dynamical evolution altered the mass function.

\section{Deriving the cumulative star cluster mass distributions from the obtained IGECMFs} \label{sect_sc_distr}

Usually, a discretized SC mass distribution is observed for SCs in or around a galaxy or a galaxy cluster and not a mass distribution function like the IGECMF. For the former, light sources are selected that presumably are SCs, and their luminosity is converted to a mass with an assumed $M$/$L$-ratio. From the individual mass estimates, a histogram is created that reveals how many SCs were found in a certain mass range. 

Therefore, we calculated the mass distributions of SCs from the obtained IGECMFs. As a byproduct, this enables us to test how elaborate the IGECMF concept is. We performed this computation in two different ways: For each SFH, we derived the cumulative distributions of SC masses, which is the number of SCs above a certain mass, $N_{\mathrm{ecl}} (> M_{\mathrm{ecl}})$, from

\begin{enumerate}
\item[\textbf{a})] the obtained IGECMFs (Sect.~\ref{sect_igecmf}),
\item[\textbf{b})] the distributions of SFRs as determined in Sect.~\ref{sect_F} (step \ref{step3} in the enumeration). 
\end{enumerate}

\noindent This calculation was restricted to the high-mass end of the SC distribution since only the most massive SCs are detected by observations. Using these two approaches, we determined the masses of the 20$\,$000 most massive SCs and compared their distributions. Ideally, both approaches should result in the same mass distributions, which would lead to the conclusion that the IGECMF correctly describes the distribution of SC masses formed during many SC formation epochs in which SC masses were distributed according to the ECMF (Eq.~\ref{ecmf}).

For \textbf{a}) the procedure is as follows: Starting at the highest $M_{\mathrm{max}}$, each of the obtained IGECMFs is integrated downward up to an individual $m_2$ until the condition in Eq.~\ref{1_igecmf} is fulfilled. The mass of the heaviest SC is determined by integrating Eq.~\ref{M_ecl_igecmf} within the same limits. According to Eq.~\ref{M_splitting}, $m_2$ is the upper integration limit of the second massive SC: Now, a limit $m_3$ is found so that Eq.~\ref{1_igecmf} holds true and Eq.~\ref{M_ecl_igecmf} is evaluated to obtain the mass of this SC. This algorithm is repeated to generate 20$\,$000 SCs in total. 

For \textbf{b}) this is done in the following way:
\begin{enumerate}
\item Determination of $M_{\mathrm{max}}$: For each SFH the highest SFR (obtained in Sect.~\ref{sect_F}, step \ref{step3} in the enumeration) is selected and converted to $M_{\mathrm{max}}$ according to Eq.~\ref{sfr}. Since the inverse function of Eq.~\ref{sfr} cannot be derived analytically, $M_{\mathrm{max}}$ is found numerically. Knowledge about $M_{\mathrm{max}}$ fully determines the ECMF (cf.~Eqs.~\ref{ecmf} and \ref{norm_k}). \label{pro1}
\item Calculating the number of SCs and their masses: With $m_1 = M_{\mathrm{max}}$, $m_2$ is evaluated from Eq.~\ref{cond1}. These integration limits are used to determine the mass of this most massive SC with Eq.~\ref{condM}. The found lower limit $m_2$ is the upper limit for the second massive SC (cf.~Eq.~\ref{M_splitting}) from which the next lower limit $m_3$ can be determined with Eq.~\ref{cond1}. Using these limits, the mass of the second massive SC is calculated (Eq.~\ref{condM}). This procedure is continued until the masses of the 20$\,$000 most massive SCs are evaluated. \label{pro2}
\item Selection of the 20$\,$000 most massive SCs: Steps~\ref{pro1} and \ref{pro2} are repeated for the second, third, fourth, and so on highest SFR. Out of all calculations, the 20$\,$000 most massive SCs are selected. 
\end{enumerate}

\begin{figure*}[t]
\includegraphics[angle=-90, width=\textwidth]{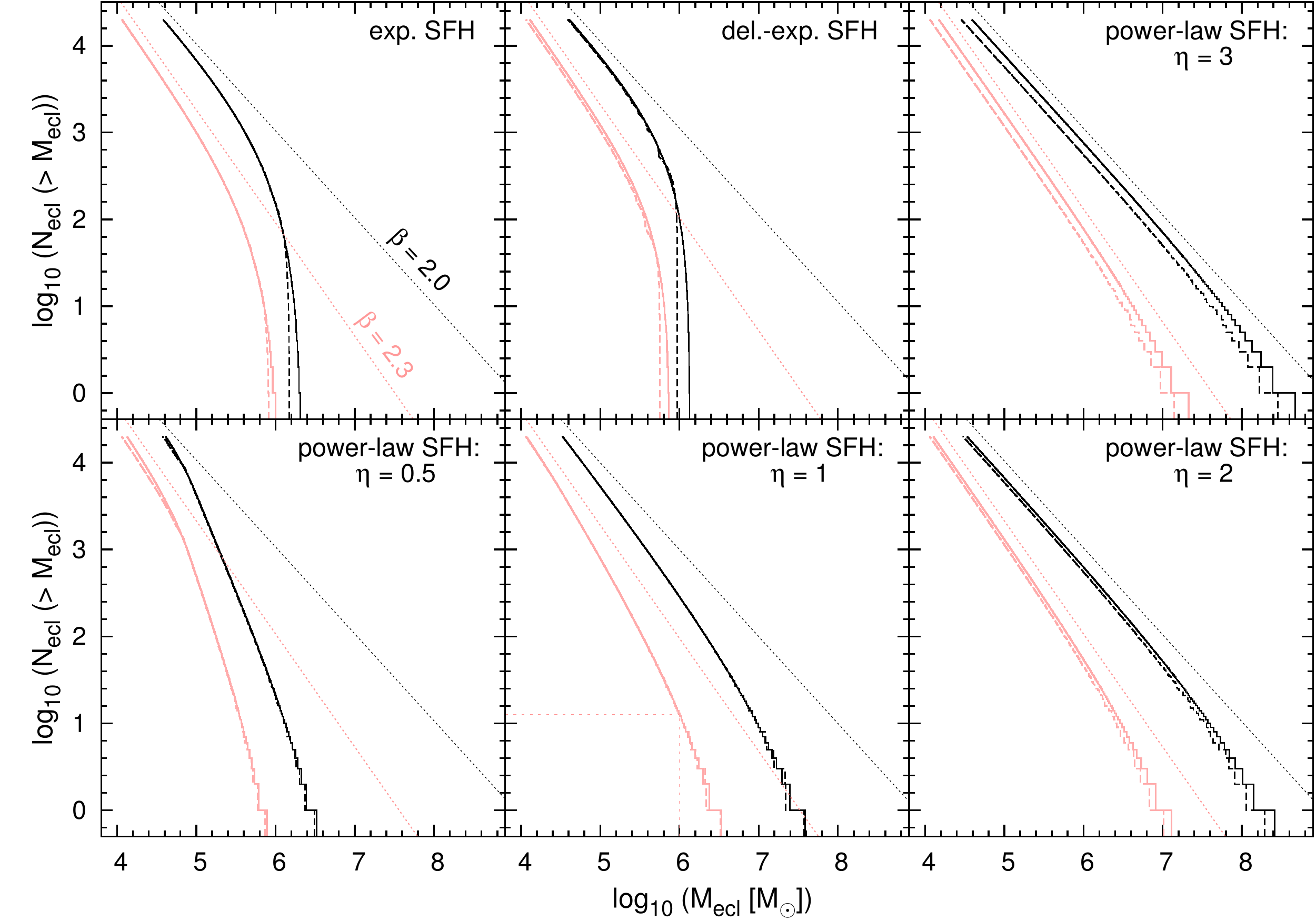}
\caption[The cumulative mass distributions of the most massive SCs for $\beta = 2.0$ and $\beta = 2.3$]{Cumulative mass distributions of the 20$\,$000 most massive SCs for the SFH as indicated in the top right corner of each panel, computed a) from the IGECMFs (continuous lines) and b) from the ECMF separately for SC formation epochs with the highest SFRs (dashed lines) for $\beta = 2.0$ (black) and $\beta = 2.3$ (red). Note the small deviations between method a) and b). For comparison, the dotted lines show the theoretical cumulative SC mass distributions for one single SC formation epoch with the same ECMF in the same color. To avoid overlaps, these lines are shifted to higher values. According to the underlying SFHs, the panels are arranged in the same manner as in Fig.~\ref{fig_F_20+23}. For instance, as indicated in the lower middle panel there are in total about $10^{1.1} \approx 13~$SCs above $M_{\mathrm{ecl}} = 10^6~\mathrm{M}_{\odot}$. (A color version of this figure is available in the online journal.)}
\label{fig_nup_20+23}
\end{figure*}

Both calculations were performed for each of the six SFHs for each $\beta$. The resulting cumulative distributions of SC masses, $N_{\mathrm{ecl}} (> M_{\mathrm{ecl}})$, are presented in black for $\beta = 2.0$ and in red for $\beta = 2.3$ in Fig.~\ref{fig_nup_20+23}. The continuous lines represent the distributions obtained from the integration of the IGECMF (a) while the dashed lines belong to those from integrating the ECMF separately for the SC formation epochs with the highest SFRs (b). The distributions are plotted such that the lines increase by unity at the mass of each SC. The dotted lines show the theoretical cumulative SC mass distributions from the ECMF with the same $\beta$ for one single SC formation epoch for comparison. Since the ECMF is assumed to be a pure power law, these mass distributions do not exhibit any steepening toward the high-mass end. To omit overlaps, they are shifted to higher values.

Two features become immediately apparent:
\begin{itemize} 
\item[\small$\bullet$] Apart from some differences (see below), the cumulative SC mass distributions obtained from a) and b) are very similar or even indistinguishable. This confirms that it almost does not matter whether one samples from the IGECMF (a) or from individual ECMFs superposed according to the SFH (b).
\item[\small$\bullet$] The resulting SC mass distributions exhibit a more or less strongly pronounced turn-down at the high-mass end, as can be seen by comparing to the theoretical SC mass distributions (dotted lines in Fig.~\ref{fig_nup_20+23}). This turn-down is explicitly visible in the SC mass distributions resulting from the exponential and the delayed-exponential SFHs as well as the power-law SFH with $\eta = 0.5$, while it is less apparent for those resulting from the power-law SFHs with $\eta > 1$. This finding is independent of $\beta$. Interestingly, a comparable steepening toward the high-mass end in the SC mass distribution has previously been detected observationally \citep{hilker09}.
\end{itemize}

The underlying $F(\mathrm{SFR})$ determines how strongly the turn-down is pronounced: If high SFRs occur frequently -- as for the exponential and particularly for the delayed-exponential SFH (Fig.~\ref{fig_F_20+23}) -- high-mass SCs are formed frequently so that the number of high-mass SCs increases fast with decreasing mass. In these cases, a turn-down in the cumulative SC mass distribution is therefore strongly pronounced; for the delayed-exponential SFH even more strongly than for the exponential SFH. In contrast, if high SFRs are rare -- as for the power-law SFHs with $\eta = 2$, and 3 -- the number of high-mass SCs increases slowly with decreasing mass, leading to a moderately well visible turn-down in the cumulative SC mass distribution. With decreasing mass, the slope of the distributions in all cases approach the slope of the theoretical cumulative SC mass distributions (dotted lines in Fig.~\ref{fig_nup_20+23}).

The main difference between the cumulative SC mass distributions is that the most massive SCs produced by a) are generally more massive than those obtained from b). The reason is the coaction between the new optimal sampling technique (Sect.~\ref{subsect_sampling}) and the difference in the shape of the IGECMF and the ECMF at the high-mass end: For a) as well as for b), to determine the most massive SC one has to start at $M_{\mathrm{max}}$, corresponding to the highest SFR, and integrate downward until the integral over the IGECMF and the ECMF, respectively, yields unity (Eqs.~\ref{1_igecmf} and \ref{cond1}). Since the IGECMF increases more strongly in integration direction than the respective ECMF, this condition will first be fulfilled for the IGECMF (a) leading to a larger $m_2$ than for the ECMF (b). Consequently, the IGECMF (a, continuous lines) produces a heavier most massive SC than the corresponding ECMF (b, dashed lines), as can be seen in Fig.~\ref{fig_nup_20+23}. The differences become smaller for increasing $\beta$ since a larger $\beta$ is connected with a steeper ECMF, which alleviates the above effect. However, for the mass distributions from the power-law SFH with $\eta = 0.5,$ the effect appears again weakly at the low-mass end.

In addition, the cumulative SC mass distributions resulting from the power-law SFHs with $\eta = 2$, and 3 from method a) are shifted to higher values over the whole mass range compared to b), whereas those from the exponential and the delayed-exponential SFH become indistinguishable with decreasing mass. This depends on the respective $F(\mathrm{SFR})$: In the case of the exponential and the delayed-exponential SFH, $F(\mathrm{SFR})$ is uniformly distributed across the SFR range (Fig.~\ref{fig_F_20+23}). Thus, relative high SFRs occur sufficiently often in these SFHs, for which reason methods a) and b) produce a similar number of high-mass SCs. As a result, method b) catches up with method a), leading to mass distributions that become equal below a certain mass. On the other hand, the power-law SFHs -- especially with $\eta = 2$, and 3 -- demonstrate that the more sparsely high SFRs are distributed (Fig.~\ref{fig_F_20+23}), the less method b) is able to catch up with method a) and the stronger the separation between the cumulative mass distributions of SCs obtained from a) and b). In contrast, methods a) and b) lead to cumulative SC mass distributions for the power-law SFHs with $\eta = 0.5$, and 1 which are nearly indistinguishable in that mass range.

According to these findings, we can summarize that the shape of the cumulative mass distributions of the 20$\,$000 most massive SCs is predominantly determined by the high-SFR end of the respective $F(\mathrm{SFR})$. The small differences in these mass distributions obtained by sampling from the IGECMF (a) and from the ECMFs determined by the highest SFRs (b) are induced by the effect discussed above, namely that at the high-mass end the IGECMF increases more strongly in integration direction than the corresponding ECMF.

In other words: The observed differences of method a) and b) demonstrate that it does matter in which order the two essential steps here -- the superposition and the discretization -- are carried out: For sampling from the ECMFs with the highest SFRs (b), first all contributing ECMFs are discretized into populations of individual SCs, and afterward, these populations are superposed. For (a), the IGECMF is first obtained from superposing all contributing ECMFs and then the IGECMF is discretized to determine the mass distribution of all SCs. As visualized, the latter leads to slightly higher masses. Nevertheless, the calculations illustrate that the cumulative SC mass distributions obtained with the two methods are very similar or even indistinguishable, leading to the conclusion that the theory of the IGECMF is a suitable concept. However, method b) is regarded to be more accurate than a) since it compiles the SC masses for each individual SC formation epoch separately and superposes them afterward, similarly to what nature does.

Interestingly, even if all IGECMFs look similar (cf.~Fig.~\ref{fig_igecmf_20+23}), the cumulative SC mass distributions are shaped differently for different SFHs, but are very similar if obtained from the same SFH (cf.~Fig.~\ref{fig_nup_20+23}). Thus, each SFH leaves through its $F(\mathrm{SFR})$ an imprint on the mass distribution of SCs independent of $\beta$. This circumstance is particularly important since it potentially provides the opportunity to deduce past star formation activities from the high-mass end of the mass distribution of an SC sample.

The derived cumulative SC mass distributions (Fig.~\ref{fig_nup_20+23}) cannot be directly compared to observed mass distributions of SCs. The reason is that the former describe hypothetical mass distribution of newly born SCs, which emerges from the superposition of SC populations that formed during many SC formation epochs. In contrast, the latter represent today's SC mass distribution, whose SC have undergone various alterations during their lifetime. The mass of an SC changes with time because of stellar evolution and the loss of stars induced by either dynamical evolution or the tidal field, which may even lead to a complete dissolution of the SC especially for low-mass ones \citep[e.g.,][]{baumgardt_makino03, lamers05, lamers10, brockamp14}. We here included none of these effects, but they will be accounted for when observational data are analyzed.

\section{Observability} \label{sect_obs}

Even if the presented model is developed to describe the mass distributions of SCs formed over a Hubble time in a galaxy cluster, the superposition principle itself is independent of the environment. Thus, this model is in principle applicable to any sample of SCs, for example,~around or in individual galaxies, and for SC formation timescales of arbitrary length. Since the superposition principle implies the development of a turn-down at the high-mass end in the mass distribution of SCs if the SFR changes with time, the question arises whether such a turn-down can be observed and if so, how reliable the interpretation of a turn-down would be. As mentioned before, we here derived the distribution of the birth stellar masses of SCs, which are different from observed SC mass functions since changes in mass during the lifetime of an SC are not taken into account.

Thus, the masses of all individual SCs in the selected sample need to be corrected for mass loss since their birth. This requires an accurate estimate of the age for each SC, which then allows computing the mass loss that is due to stellar and internal dynamical evolution. According to their age, the upper mass limit needs to be calculated below which SCs are expected to have dissolved since their birth. It is difficult to accurately estimate the number and mass distribution of completely dissolved or disrupted SCs -- for instance, by infant mortality -- therefore it is necessary to set a lower mass limit at the high-mass end down to which SCs in the sample are considered. This mass limit should be well above the dissolution limit to avoid the difficult mass correction for already dissolved SCs and thus the introduction of substantial uncertainties at the lower mass limit of the considered SC sample. The observational completeness limit might restrict the allowed SC mass range even more strongly, while the availability of more parameters such as the metallicity or the abundance of different elements might give further constraints on the age and the mass-loss history.

In addition to stellar and internal dynamical evolution, each SC will suffer from mass loss that is due to the gravitational pull of the tidal field during its orbit. One can estimate how strong the tidal field acted on the individual SC and how much mass was lost since its birth if the position of each SC and its orbit is known. However, the actual mass loss in a tidal field highly depends on the geometry of the orbit \citep[e.g.,][]{baumgardt_makino03}, which is usually unknown and very difficult to determine. In this case, one could at least estimate for a statistical distribution of orbits how the SC mass function as a whole might have changed with time. This will lead to an increasing statistical uncertainty in the mass distribution. Carrying out these corrections for the SCs in the sample finally leads to the mass distribution of their birth stellar masses.

However, before comparing the corrected SC mass distribution with theoretical mass distributions by using the method we presented here, we have to take into account that the theoretical mass distributions were generated by using the introduced new sampling technique, which distributes SC masses exactly according to the underlying mass function to avoid introducing Poisson noise. In contrast, the observed SC mass distribution carries at least some statistical uncertainties that depend on the accuracy of the mass correction. Thus, a range of matching SFR distribution functions, $F(\mathrm{SFR})$, is expected rather than one well-defined $F(\mathrm{SFR})$. However, the reliability of the analysis can be enhanced by a prospective sample selection: Either the selected SCs are of similar age since in this case one expects that all effects leading to mass loss act on those SCs in a similar way. Alternatively, the selected sample comprises a large number of SCs, which reduces the relative statistical uncertainties. Apparently, the best would be to combine both aspects. Conceivable applications are listed below.

\paragraph{SC formation history of the MW}\mbox{}\\
The MW, as many other galaxies today, is rather in a quiescent mode of SC formation. Thus, recent SFRs are usually moderate, leading to the formation of SCs that only extend to intermediate masses -- unless the SC formation activity is pushed as a result of~an interaction with another galaxy, for instance. The MW has an SFR of about $2~\mathrm{M}_{\odot}/\rm{yr}$ \citep[e.g.,][see also references therein]{chomiuk11}, while the most massive young SCs typically have masses below $10^5~\mathrm{M}_{\odot}$ \citep[e.g.,][]{figer06, mengel07, davies07}. However, there are different estimates for the lifetimes of such SCs: \citet{lamers05, lamers05b} expected from their semi-analytical model an SC with a mass of $10^4~\mathrm{M}_{\odot}$ to dissolve in about 1~Gyr, while collisional $N$-body computations of \citet{baumgardt_makino03} showed that an SC with $\approx 3 \times 10^4~\mathrm{M}_{\odot}$ at solar position in the MW will survive about a Hubble time. Apparently, these two estimates alone cannot pin down how far in the past the SC formation history can be constrained using the most massive SCs in the MW. Moreover, this example shows that due to a limited total number of SCs available for the analysis, no very robust result for the SC formation history is to be expected.

\paragraph{SC formation history of a galaxy cluster}\mbox{}\\
As indicated above, the optimal SC sample should comprise a large number of objects of similar age. Such a SC sample can be found in galaxy clusters like the Fornax galaxy cluster: It is expected that numerous SCs and in particular very massive ones were formed in interactions of individual galaxies during the early assembly of the host galaxy cluster. At least the most massive SCs are able to survive over a Hubble time -- and are observed today as GCs and partly as UCDs, most of them having an age of the order 10~Gyr \citep[e.g.,][]{forbes01, kundu05, hempel07, chilingarian11, francis12}. Fortunately, the very massive SCs are least affected by changes in their mass during their lifetime compared to low-mass SCs \citep[e.g.,][]{baumgardt_makino03}. Thus, it is expected that a turn-down in the mass function, if it exists after the birth of those SCs, should be preserved. This is indeed the case: \citet[his Fig.~4,]{hilker09} reported a steepening of the mass function toward the high-mass end, where about 1$\,$000 SC-like objects are found with masses above $10^6~\mathrm{M}_{\odot}$ (M. Hilker, private communication). Most of them are GCs, while the heaviest objects are UCDs, which are compatible with being very massive GCs \citep{mieske12}. This demonstrates that such rich GC systems represent an ideal, statistically significant data set for the application of the introduced superposition principle with which the formation history of these GCs can be uncovered.

\section{Conclusions} \label{sect_concl}

We investigated how the masses of SCs are distributed for different SFHs in a galaxy cluster environment by superposition of all newly born SC populations of all SC formation epochs. This mass function is composed of the SCs' birth stellar masses since no evolutionary changes of the mass function are taken into account. While deriving this mass function, we developed the following:

\begin{itemize}
\item[\small$\bullet$] A new analytical sampling technique without stochastic fluctuations was introduced as an improved optimal sampling method. This technique enables accurately extracting the number of SCs as well as their individual masses from the underlying ECMF at the same time (Sect.~\ref{subsect_sampling}).
\item[\small$\bullet$] The length of one star formation epoch, $\delta t$, that means, the time it takes to form an SC population out of the interstellar medium so that the ECMF is fully populated (Sect.~\ref{subsect_sampling}), was determined as a function of the index $\beta$ using the SFR-$M_{\mathrm{ecl,max}}$ relation (Eq.~\ref{sfr_M_eclmax}, Fig.~\ref{fig_sfr_Mmax5}). For $2.0 \le \beta \le 2.3,$ we obtained $2~\mathrm{Myr} \lesssim \delta t \lesssim 20~\mathrm{Myr,}$ in agreement with observational constraints (Sect.~\ref{sect_dt}).
\end{itemize}

Six SFHs (five decreasing and one increasing in the beginning and decreasing afterward, Sect.~\ref{sect_SFH}) were considered, which led to the formation of SCs of $10^{10}~\mathrm{M}_{\odot}$ in total. Each SFH was divided into formation epochs of length $\delta t$ (Fig.~\ref{fig_sketch}), and we determined for each SC formation epoch its corresponding SFR. From the distribution of SFRs, we computed for each SFH the distribution function of SFRs, $F(\mathrm{SFR})$, which carries information about the past star formation activities (Sect.~\ref{sect_F}):

\begin{itemize}
\item[\small$\bullet$] For all monotonically declining SFHs used here, we found that the corresponding $F(\mathrm{SFR})$ was a pure power law (Fig.~\ref{fig_F_20+23}). This finding is independent of any input parameter and was also derived analytically.
\end{itemize}

\noindent Based on $F(\mathrm{SFR})$, we evaluated the IGECMF.
This represents how the birth stellar masses of all SCs is distributed for each SFH (Sect.~\ref{sect_igecmf}). Remarkably, we found that 

\begin{itemize}
\item[\small$\bullet$] all IGECMFs steepen toward the high-mass end (Fig.~\ref{fig_igecmf_20+23}). This finding is independent of any parameter such as the parental ECMF or the underlying SFH.
\item[\small$\bullet$] the appearance of a turn-down at the high-mass end is caused by an SFH that changes with time.
\end{itemize}

\noindent If the SFR changes with time, the mass of the most massive SC, $M_{\mathrm{ecl,max}}$, also changes according to the SFR-$M_{\mathrm{ecl,max}}$ relation (Eq.~\ref{sfr_M_eclmax}, Fig.~\ref{fig_sfr_Mmax5}). Thus, the ECMF is populated up to a variable upper limit, while the low-mass end is filled with SCs as before. Consequently, there will be fewer very massive SCs per low-mass SC, for which reason the IGECMF falls off more steeply than the parental ECMF. On the other hand, if the SFR is constant over several SC formation epochs, no turn-down in the IGECMF is expected.

For each IGECMF -- determined by the index of the underlying ECMF and the SFH -- we computed the cumulative mass distribution of SCs using the new sampling technique (Sect.~\ref{sect_sc_distr}). The turn-down of the IGECMF at the high-mass end causes fewer very massive SCs to be produced per low-mass SC than for an extrapolated ECMF with the same slope. Thus, the cumulative mass distributions of SCs also steepen toward the high-mass end (Fig.~\ref{fig_nup_20+23}). The shape of the high-mass end even differs among the cumulative SC mass distributions: How strongly the steepening toward higher masses is pronounced essentially depends on the underlying SFH through its $F(\mathrm{SFR})$. {From this, we conclude:} 
 
\begin{itemize}
\item[\small$\bullet$] Past star formation activities leave an imprint on an SC mass distribution: This imprint is the shape of the mass function at the high-mass end.
\end{itemize}

We summarize our key conclusion as follows:

\begin{itemize}
\item[\small$\bullet$] It is expected that a mass function composed of a superposition of SC populations, each described by a truncated power law and fully populated during a formation epoch of length $\delta t,$ will steepen toward the high-mass end if the SFR changed during an SC formation period longer than $\delta t$.
\end{itemize}

The appearance of a turn-down in an SC mass function allows drawing a substantial conclusion: it might reveal why the cluster initial mass function (CIMF) has a Schechter-type behavior \citep{schechter76} at the high-mass end, which has been detected by various authors \citep{gieles06letter, bastian08, larsen09, vansevicius09, bastian12} and was used to describe the distribution of SC masses \citep[e.g.,][]{gieles09}. According to our examination, the turn-down can be explained by a superposition of several SC populations that formed during different formation epochs at different SFRs. If this is indeed the case, then the CIMF can be interpreted as a local IGECMF since the superposition principle will work independent of the environment. Consequently, it is not surprising that a Schechter-type function offers a better description than a pure power law.

Additionally, given our findings, it appears natural that the exponential truncation mass, $M_{\mathrm{c}}$ or $M_{\star}$, varies among different SC ensembles and environments (cf.~\citealt{gieles06, gieles06letter, bastian08, larsen09, bastian12}; see also \citealt{portegies_zwart10}): Using the SFR-$M_{\mathrm{ecl,max}}$ relation, the mass range containing the turn-down can be converted into an SFR range to determine how strongly the SFR changed in the past during the formation of the considered SC ensemble -- which will be unique for each environment.

A turn-down in the SC mass distribution is only expected if the SFR changes with time. If SCs are formed over several SC formation epochs at a constant SFR, the power-law behavior of the SC mass function will remain unchanged. If the mass function of an SC ensemble does not show any tendency of a turn-down, this then indicates that the SCs have formed at a more or less constant SFR.

In addition to the Schechter-type behavior of the CIMF, a turn-down or bend in the mass distribution was observed by \citet{hilker09} for GCs and UCDs around major galaxies in galaxy clusters. Our investigations provide a natural explanation for the observed steepening: These objects could simply have formed during different formation epochs at different SFRs. This is a quite reasonable assumption since major galaxies in galaxy clusters could indeed have had several interactions with infalling galaxies, during which SCs at different SFRs were formed. Thus, the superposition principle offers a reasonable physical explanation for the appearance of the turn-down.

It is currently only poorly known under which conditions the GCs and UCDs formed. One method for deriving the formation history of an SC ensemble has been described in \citet{maschberger07} and relies on the information about the mass and in particular the age of each SC. Unfortunately, it is not promising to apply this method to a sample of GCs and UCDs because these objects are assumed to be very old, and their age estimates suffer from large uncertainties. 

Our investigations provide an alternative approach to derive the formation history of an SC ensemble without the requirement for detailed age information: Our idea is to reverse the direction of thought according to the main finding of our analysis: to use the high-mass end of a known GC/UCD mass distribution to deduce which SFRs are necessary to reproduce the observed mass distribution. This ansatz is considered to be promising since it requires examining the shape of the high-mass end where the most massive SCs are located, which are least affected by dynamical evolution and have the highest probability of surviving violent interactions with the host galaxy. Before doing that, the observed GC/UCD mass distribution has to be corrected for all possible evolutionary effects that lead to changes in the mass distribution. This ansatz will be used in a future paper to derive the necessary distribution of SFRs from which $F(\mathrm{SFR})$ and possibly also the SFH can be deduced. Hopefully, this will shed light on the way major galaxies in galaxy clusters assembled and under which conditions the surrounding GC/UCD systems formed.

\bibliography{d3-aa}   

\begin{thebibliography}{85}
\expandafter\ifx\csname natexlab\endcsname\relax\def\natexlab#1{#1}\fi

\bibitem[{{Adamo} {et~al.}(2011){Adamo}, {{\"O}stlin}, \&
  {Zackrisson}}]{adamo11}
{Adamo}, A., {{\"O}stlin}, G., \& {Zackrisson}, E. 2011, \mnras, 417, 1904

\bibitem[{{Alexander} \& {Gieles}(2012)}]{alexander12}
{Alexander}, P.~E.~R. \& {Gieles}, M. 2012, \mnras, 422, 3415

\bibitem[{{Alexander} {et~al.}(2014){Alexander}, {Gieles}, {Lamers}, \&
  {Baumgardt}}]{alexander14}
{Alexander}, P.~E.~R., {Gieles}, M., {Lamers}, H.~J.~G.~L.~M., \& {Baumgardt},
  H. 2014, \mnras, 442, 1265

\bibitem[{{Bastian}(2008)}]{bastian08}
{Bastian}, N. 2008, \mnras, 390, 759

\bibitem[{{Bastian} {et~al.}(2012{\natexlab{a}}){Bastian}, {Adamo}, {Gieles},
  {Silva-Villa}, {Lamers}, {Larsen}, {Smith}, {Konstantopoulos}, \&
  {Zackrisson}}]{bastian12}
{Bastian}, N., {Adamo}, A., {Gieles}, M., {et~al.} 2012{\natexlab{a}}, \mnras,
  419, 2606

\bibitem[{{Bastian} {et~al.}(2012{\natexlab{b}}){Bastian}, {Konstantopoulos},
  {Trancho}, {Weisz}, {Larsen}, {Fouesneau}, {Kaschinski}, \&
  {Gieles}}]{bastian12rn}
{Bastian}, N., {Konstantopoulos}, I.~S., {Trancho}, G., {et~al.}
  2012{\natexlab{b}}, \aap, 541, A25

\bibitem[{{Baumgardt} \& {Makino}(2003)}]{baumgardt_makino03}
{Baumgardt}, H. \& {Makino}, J. 2003, \mnras, 340, 227

\bibitem[{{Beasley} {et~al.}(2008){Beasley}, {Bridges}, {Peng}, {Harris},
  {Harris}, {Forbes}, \& {Mackie}}]{beasley08}
{Beasley}, M.~A., {Bridges}, T., {Peng}, E., {et~al.} 2008, \mnras, 386, 1443

\bibitem[{{Bik} {et~al.}(2003){Bik}, {Lamers}, {Bastian}, {Panagia}, \&
  {Romaniello}}]{bik03}
{Bik}, A., {Lamers}, H.~J.~G.~L.~M., {Bastian}, N., {Panagia}, N., \&
  {Romaniello}, M. 2003, \aap, 397, 473

\bibitem[{{Billett} {et~al.}(2002){Billett}, {Hunter}, \&
  {Elmegreen}}]{billett02}
{Billett}, O.~H., {Hunter}, D.~A., \& {Elmegreen}, B.~G. 2002, \aj, 123, 1454

\bibitem[{{Bonatto} \& {Bica}(2012)}]{bonatto12}
{Bonatto}, C. \& {Bica}, E. 2012, \mnras, 2898

\bibitem[{{Bonnell} {et~al.}(2006){Bonnell}, {Dobbs}, {Robitaille}, \&
  {Pringle}}]{bonnell06}
{Bonnell}, I.~A., {Dobbs}, C.~L., {Robitaille}, T.~P., \& {Pringle}, J.~E.
  2006, \mnras, 365, 37

\bibitem[{{Brockamp} {et~al.}(2014){Brockamp}, {K{\"u}pper}, {Thies},
  {Baumgardt}, \& {Kroupa}}]{brockamp14}
{Brockamp}, M., {K{\"u}pper}, A.~H.~W., {Thies}, I., {Baumgardt}, H., \&
  {Kroupa}, P. 2014, \mnras, 441, 150

\bibitem[{{Chandar} {et~al.}(2011){Chandar}, {Whitmore}, {Calzetti}, {Di Nino},
  {Kennicutt}, {Regan}, \& {Schinnerer}}]{chandar11}
{Chandar}, R., {Whitmore}, B.~C., {Calzetti}, D., {et~al.} 2011, \apj, 727, 88

\bibitem[{{Chandar} {et~al.}(2010){Chandar}, {Whitmore}, {Kim}, {Kaleida},
  {Mutchler}, {Calzetti}, {Saha}, {O'Connell}, {Balick}, {Bond}, {Carollo},
  {Disney}, {Dopita}, {Frogel}, {Hall}, {Holtzman}, {Kimble}, {McCarthy},
  {Paresce}, {Silk}, {Trauger}, {Walker}, {Windhorst}, \& {Young}}]{chandar10}
{Chandar}, R., {Whitmore}, B.~C., {Kim}, H., {et~al.} 2010, \apj, 719, 966

\bibitem[{{Chilingarian} {et~al.}(2011){Chilingarian}, {Mieske}, {Hilker}, \&
  {Infante}}]{chilingarian11}
{Chilingarian}, I.~V., {Mieske}, S., {Hilker}, M., \& {Infante}, L. 2011,
  \mnras, 412, 1627

\bibitem[{{Chomiuk} \& {Povich}(2011)}]{chomiuk11}
{Chomiuk}, L. \& {Povich}, M.~S. 2011, \aj, 142, 197

\bibitem[{{C{\^o}t{\'e}} {et~al.}(1998){C{\^o}t{\'e}}, {Marzke}, \&
  {West}}]{cote98}
{C{\^o}t{\'e}}, P., {Marzke}, R.~O., \& {West}, M.~J. 1998, \apj, 501, 554

\bibitem[{{Daddi} {et~al.}(2007){Daddi}, {Dickinson}, {Morrison}, {Chary},
  {Cimatti}, {Elbaz}, {Frayer}, {Renzini}, {Pope}, {Alexander}, {Bauer},
  {Giavalisco}, {Huynh}, {Kurk}, \& {Mignoli}}]{daddi07}
{Daddi}, E., {Dickinson}, M., {Morrison}, G., {et~al.} 2007, \apj, 670, 156

\bibitem[{{Davies} {et~al.}(2007){Davies}, {Figer}, {Kudritzki}, {MacKenty},
  {Najarro}, \& {Herrero}}]{davies07}
{Davies}, B., {Figer}, D.~F., {Kudritzki}, R.-P., {et~al.} 2007, \apj, 671, 781

\bibitem[{{de Grijs} \& {Anders}(2006)}]{degrijs_anders06}
{de Grijs}, R. \& {Anders}, P. 2006, \mnras, 366, 295

\bibitem[{{de Grijs} {et~al.}(2003){de Grijs}, {Anders}, {Bastian}, {Lynds},
  {Lamers}, \& {O'Neil}}]{degrijs03}
{de Grijs}, R., {Anders}, P., {Bastian}, N., {et~al.} 2003, \mnras, 343, 1285

\bibitem[{{de Grijs} \& {Goodwin}(2008)}]{degrijs_goodwin08}
{de Grijs}, R. \& {Goodwin}, S.~P. 2008, \mnras, 383, 1000

\bibitem[{{Dowell} {et~al.}(2008){Dowell}, {Buckalew}, \& {Tan}}]{dowell08}
{Dowell}, J.~D., {Buckalew}, B.~A., \& {Tan}, J.~C. 2008, \aj, 135, 823

\bibitem[{{Egusa} {et~al.}(2009){Egusa}, {Kohno}, {Sofue}, {Nakanishi}, \&
  {Komugi}}]{egusa09}
{Egusa}, F., {Kohno}, K., {Sofue}, Y., {Nakanishi}, H., \& {Komugi}, S. 2009,
  \apj, 697, 1870

\bibitem[{{Egusa} {et~al.}(2004){Egusa}, {Sofue}, \& {Nakanishi}}]{egusa04}
{Egusa}, F., {Sofue}, Y., \& {Nakanishi}, H. 2004, \pasj, 56, L45

\bibitem[{{Fall}(2004)}]{fall04asp}
{Fall}, S.~M. 2004, in Astronomical Society of the Pacific Conference Series,
  Vol. 322, The Formation and Evolution of Massive Young Star Clusters, ed.
  H.~J.~G.~L.~M. {Lamers}, L.~J. {Smith}, \& A.~{Nota}, 399

\bibitem[{{Figer} {et~al.}(2006){Figer}, {MacKenty}, {Robberto}, {Smith},
  {Najarro}, {Kudritzki}, \& {Herrero}}]{figer06}
{Figer}, D.~F., {MacKenty}, J.~W., {Robberto}, M., {et~al.} 2006, \apj, 643,
  1166

\bibitem[{{Forbes} {et~al.}(2001){Forbes}, {Beasley}, {Brodie}, \&
  {Kissler-Patig}}]{forbes01}
{Forbes}, D.~A., {Beasley}, M.~A., {Brodie}, J.~P., \& {Kissler-Patig}, M.
  2001, \apjl, 563, L143

\bibitem[{{Francis} {et~al.}(2012){Francis}, {Drinkwater}, {Chilingarian},
  {Bolt}, \& {Firth}}]{francis12}
{Francis}, K.~J., {Drinkwater}, M.~J., {Chilingarian}, I.~V., {Bolt}, A.~M., \&
  {Firth}, P. 2012, \mnras, 425, 325

\bibitem[{{Fukui} {et~al.}(1999){Fukui}, {Mizuno}, {Yamaguchi}, {Mizuno},
  {Onishi}, {Ogawa}, {Yonekura}, {Kawamura}, {Tachihara}, {Xiao}, {Yamaguchi},
  {Hara}, {Hayakawa}, {Kato}, {Abe}, {Saito}, {Mano}, {Matsunaga}, {Mine},
  {Moriguchi}, {Aoyama}, {Asayama}, {Yoshikawa}, \& {Rubio}}]{fukui99}
{Fukui}, Y., {Mizuno}, N., {Yamaguchi}, R., {et~al.} 1999, \pasj, 51, 745

\bibitem[{{Gavazzi} {et~al.}(2002){Gavazzi}, {Bonfanti}, {Sanvito}, {Boselli},
  \& {Scodeggio}}]{gavazzi02}
{Gavazzi}, G., {Bonfanti}, C., {Sanvito}, G., {Boselli}, A., \& {Scodeggio}, M.
  2002, \apj, 576, 135

\bibitem[{{Georgiev} {et~al.}(2012){Georgiev}, {Goudfrooij}, \&
  {Puzia}}]{georgiev12}
{Georgiev}, I.~Y., {Goudfrooij}, P., \& {Puzia}, T.~H. 2012, \mnras, 420, 1317

\bibitem[{{Gieles}(2009)}]{gieles09}
{Gieles}, M. 2009, \mnras, 394, 2113

\bibitem[{{Gieles} {et~al.}(2014){Gieles}, {Alexander}, {Lamers}, \&
  {Baumgardt}}]{gieles14}
{Gieles}, M., {Alexander}, P.~E.~R., {Lamers}, H.~J.~G.~L.~M., \& {Baumgardt},
  H. 2014, \mnras, 437, 916

\bibitem[{{Gieles} {et~al.}(2011){Gieles}, {Heggie}, \& {Zhao}}]{gieles11}
{Gieles}, M., {Heggie}, D.~C., \& {Zhao}, H. 2011, \mnras, 413, 2509

\bibitem[{{Gieles} {et~al.}(2006{\natexlab{a}}){Gieles}, {Larsen}, {Bastian},
  \& {Stein}}]{gieles06}
{Gieles}, M., {Larsen}, S.~S., {Bastian}, N., \& {Stein}, I.~T.
  2006{\natexlab{a}}, \aap, 450, 129

\bibitem[{{Gieles} {et~al.}(2006{\natexlab{b}}){Gieles}, {Larsen},
  {Scheepmaker}, {Bastian}, {Haas}, \& {Lamers}}]{gieles06letter}
{Gieles}, M., {Larsen}, S.~S., {Scheepmaker}, R.~A., {et~al.}
  2006{\natexlab{b}}, \aap, 446, L9

\bibitem[{{Haas} \& {Anders}(2010)}]{haas_anders10}
{Haas}, M.~R. \& {Anders}, P. 2010, \aap, 512, A79

\bibitem[{{Hempel} {et~al.}(2007){Hempel}, {Kissler-Patig}, {Puzia}, \&
  {Hilker}}]{hempel07}
{Hempel}, M., {Kissler-Patig}, M., {Puzia}, T.~H., \& {Hilker}, M. 2007, \aap,
  463, 493

\bibitem[{{Hilker}(2009)}]{hilker09}
{Hilker}, M. 2009, ArXiv e-prints

\bibitem[{{Hunter} {et~al.}(2003){Hunter}, {Elmegreen}, {Dupuy}, \&
  {Mortonson}}]{hunter03}
{Hunter}, D.~A., {Elmegreen}, B.~G., {Dupuy}, T.~J., \& {Mortonson}, M. 2003,
  \aj, 126, 1836

\bibitem[{{Kroupa}(2002)}]{kroupa02}
{Kroupa}, P. 2002, \mnras, 330, 707

\bibitem[{{Kroupa}(2015)}]{kroupa15review}
{Kroupa}, P. 2015, Canadian Journal of Physics, 93, 169

\bibitem[{{Kroupa} \& {Boily}(2002)}]{kroupa_boily02}
{Kroupa}, P. \& {Boily}, C.~M. 2002, \mnras, 336, 1188

\bibitem[{{Kroupa} \& {Weidner}(2003)}]{kroupa_weidner03}
{Kroupa}, P. \& {Weidner}, C. 2003, \apj, 598, 1076

\bibitem[{{Kroupa} {et~al.}(2013){Kroupa}, {Weidner}, {Pflamm-Altenburg},
  {Thies}, {Dabringhausen}, {Marks}, \& {Maschberger}}]{kroupa13rev}
{Kroupa}, P., {Weidner}, C., {Pflamm-Altenburg}, J., {et~al.} 2013, {The
  Stellar and Sub-Stellar Initial Mass Function of Simple and Composite
  Populations} ({Oswalt}, T.~D. and {Gilmore}, G.), 115

\bibitem[{{Kundu} {et~al.}(2005){Kundu}, {Zepf}, {Hempel}, {Morton}, {Ashman},
  {Maccarone}, {Kissler-Patig}, {Puzia}, \& {Vesperini}}]{kundu05}
{Kundu}, A., {Zepf}, S.~E., {Hempel}, M., {et~al.} 2005, \apjl, 634, L41

\bibitem[{{Lada} \& {Lada}(2003)}]{lada_lada03}
{Lada}, C.~J. \& {Lada}, E.~A. 2003, \araa, 41, 57

\bibitem[{{Lamers} {et~al.}(2010){Lamers}, {Baumgardt}, \& {Gieles}}]{lamers10}
{Lamers}, H.~J.~G.~L.~M., {Baumgardt}, H., \& {Gieles}, M. 2010, \mnras, 409,
  305

\bibitem[{{Lamers} {et~al.}(2013){Lamers}, {Baumgardt}, \& {Gieles}}]{lamers13}
{Lamers}, H.~J.~G.~L.~M., {Baumgardt}, H., \& {Gieles}, M. 2013, \mnras, 433,
  1378

\bibitem[{{Lamers} {et~al.}(2005{\natexlab{a}}){Lamers}, {Gieles}, {Bastian},
  {Baumgardt}, {Kharchenko}, \& {Portegies Zwart}}]{lamers05}
{Lamers}, H.~J.~G.~L.~M., {Gieles}, M., {Bastian}, N., {et~al.}
  2005{\natexlab{a}}, \aap, 441, 117

\bibitem[{{Lamers} {et~al.}(2005{\natexlab{b}}){Lamers}, {Gieles}, \&
  {Portegies Zwart}}]{lamers05b}
{Lamers}, H.~J.~G.~L.~M., {Gieles}, M., \& {Portegies Zwart}, S.~F.
  2005{\natexlab{b}}, \aap, 429, 173

\bibitem[{{Larsen}(2002)}]{larsen02}
{Larsen}, S.~S. 2002, \aj, 124, 1393

\bibitem[{{Larsen}(2009)}]{larsen09}
{Larsen}, S.~S. 2009, \aap, 494, 539

\bibitem[{{Larsen} \& {Richtler}(2000)}]{larsen_richtler00}
{Larsen}, S.~S. \& {Richtler}, T. 2000, \aap, 354, 836

\bibitem[{{Le Floc'h} {et~al.}(2005){Le Floc'h}, {Papovich}, {Dole}, {Bell},
  {Lagache}, {Rieke}, {Egami}, {P{\'e}rez-Gonz{\'a}lez}, {Alonso-Herrero},
  {Rieke}, {Blaylock}, {Engelbracht}, {Gordon}, {Hines}, {Misselt}, {Morrison},
  \& {Mould}}]{lefloch05}
{Le Floc'h}, E., {Papovich}, C., {Dole}, H., {et~al.} 2005, \apj, 632, 169

\bibitem[{{Lilly} {et~al.}(1996){Lilly}, {Le Fevre}, {Hammer}, \&
  {Crampton}}]{lilly96}
{Lilly}, S.~J., {Le Fevre}, O., {Hammer}, F., \& {Crampton}, D. 1996, \apjl,
  460, L1

\bibitem[{{Madau} {et~al.}(1998){Madau}, {Pozzetti}, \& {Dickinson}}]{madau98}
{Madau}, P., {Pozzetti}, L., \& {Dickinson}, M. 1998, \apj, 498, 106

\bibitem[{{Maschberger} \& {Kroupa}(2007)}]{maschberger07}
{Maschberger}, T. \& {Kroupa}, P. 2007, \mnras, 379, 34

\bibitem[{{McCrady} \& {Graham}(2007)}]{mccrady_graham07}
{McCrady}, N. \& {Graham}, J.~R. 2007, \apj, 663, 844

\bibitem[{{Mengel} \& {Tacconi-Garman}(2007)}]{mengel07}
{Mengel}, S. \& {Tacconi-Garman}, L.~E. 2007, \aap, 466, 151

\bibitem[{{Mieske} {et~al.}(2012){Mieske}, {Hilker}, \& {Misgeld}}]{mieske12}
{Mieske}, S., {Hilker}, M., \& {Misgeld}, I. 2012, \aap, 537, A3

\bibitem[{{Norris} \& {Kannappan}(2011)}]{norris_kannappan11}
{Norris}, M.~A. \& {Kannappan}, S.~J. 2011, \mnras, 414, 739

\bibitem[{{Pflamm-Altenburg} {et~al.}(2013){Pflamm-Altenburg},
  {Gonz{\'a}lez-L{\'o}pezlira}, \& {Kroupa}}]{pflamm-altenburg13}
{Pflamm-Altenburg}, J., {Gonz{\'a}lez-L{\'o}pezlira}, R.~A., \& {Kroupa}, P.
  2013, \mnras, 435, 2604

\bibitem[{{Pflamm-Altenburg} \& {Kroupa}(2006)}]{pflamm-altenburg06}
{Pflamm-Altenburg}, J. \& {Kroupa}, P. 2006, \mnras, 373, 295

\bibitem[{{Pflamm-Altenburg} {et~al.}(2007){Pflamm-Altenburg}, {Weidner}, \&
  {Kroupa}}]{pflamm-altenburg07}
{Pflamm-Altenburg}, J., {Weidner}, C., \& {Kroupa}, P. 2007, \apj, 671, 1550

\bibitem[{{Portegies Zwart} {et~al.}(2010){Portegies Zwart}, {McMillan}, \&
  {Gieles}}]{portegies_zwart10}
{Portegies Zwart}, S.~F., {McMillan}, S.~L.~W., \& {Gieles}, M. 2010, \araa,
  48, 431

\bibitem[{{Randriamanakoto} {et~al.}(2013){Randriamanakoto}, {Escala},
  {V{\"a}is{\"a}nen}, {Kankare}, {Kotilainen}, {Mattila}, \&
  {Ryder}}]{randria13}
{Randriamanakoto}, Z., {Escala}, A., {V{\"a}is{\"a}nen}, P., {et~al.} 2013,
  \apjl, 775, L38

\bibitem[{{Rosolowsky} {et~al.}(2007){Rosolowsky}, {Keto}, {Matsushita}, \&
  {Willner}}]{rosolowsky07}
{Rosolowsky}, E., {Keto}, E., {Matsushita}, S., \& {Willner}, S.~P. 2007, \apj,
  661, 830

\bibitem[{{Sandage}(1986)}]{sandage86}
{Sandage}, A. 1986, \aap, 161, 89

\bibitem[{{Schechter}(1976)}]{schechter76}
{Schechter}, P. 1976, \apj, 203, 297

\bibitem[{{Schiminovich} {et~al.}(2005){Schiminovich}, {Ilbert}, {Arnouts},
  {Milliard}, {Tresse}, {Le F{\`e}vre}, {Treyer}, {Wyder}, {Budav{\'a}ri},
  {Zucca}, {Zamorani}, {Martin}, {Adami}, {Arnaboldi}, {Bardelli}, {Barlow},
  {Bianchi}, {Bolzonella}, {Bottini}, {Byun}, {Cappi}, {Contini}, {Charlot},
  {Donas}, {Forster}, {Foucaud}, {Franzetti}, {Friedman}, {Garilli},
  {Gavignaud}, {Guzzo}, {Heckman}, {Hoopes}, {Iovino}, {Jelinsky}, {Le Brun},
  {Lee}, {Maccagni}, {Madore}, {Malina}, {Marano}, {Marinoni}, {McCracken},
  {Mazure}, {Meneux}, {Morrissey}, {Neff}, {Paltani}, {Pell{\`o}}, {Picat},
  {Pollo}, {Pozzetti}, {Radovich}, {Rich}, {Scaramella}, {Scodeggio},
  {Seibert}, {Siegmund}, {Small}, {Szalay}, {Vettolani}, {Welsh}, {Xu}, \&
  {Zanichelli}}]{schiminovich05}
{Schiminovich}, D., {Ilbert}, O., {Arnouts}, S., {et~al.} 2005, \apjl, 619, L47

\bibitem[{{Tamburro} {et~al.}(2008){Tamburro}, {Rix}, {Walter}, {Brinks}, {de
  Blok}, {Kennicutt}, \& {Mac Low}}]{tamburro08}
{Tamburro}, D., {Rix}, H.-W., {Walter}, F., {et~al.} 2008, \aj, 136, 2872

\bibitem[{{Vansevi{\v c}ius} {et~al.}(2009){Vansevi{\v c}ius}, {Kodaira},
  {Narbutis}, {Stonkut{\.e}}, {Brid{\v z}ius}, {Deveikis}, \&
  {Semionov}}]{vansevicius09}
{Vansevi{\v c}ius}, V., {Kodaira}, K., {Narbutis}, D., {et~al.} 2009, \apj,
  703, 1872

\bibitem[{{Weidner} {et~al.}(2010{\natexlab{a}}){Weidner}, {Bonnell}, \&
  {Zinnecker}}]{weidner10}
{Weidner}, C., {Bonnell}, I.~A., \& {Zinnecker}, H. 2010{\natexlab{a}}, \apj,
  724, 1503

\bibitem[{{Weidner} \& {Kroupa}(2004)}]{weidner_kroupa04}
{Weidner}, C. \& {Kroupa}, P. 2004, \mnras, 348, 187

\bibitem[{{Weidner} \& {Kroupa}(2005)}]{weidner_kroupa05}
{Weidner}, C. \& {Kroupa}, P. 2005, \apj, 625, 754

\bibitem[{{Weidner} {et~al.}(2010{\natexlab{b}}){Weidner}, {Kroupa}, \&
  {Bonnell}}]{weidner_k_b10}
{Weidner}, C., {Kroupa}, P., \& {Bonnell}, I.~A.~D. 2010{\natexlab{b}}, \mnras,
  401, 275

\bibitem[{{Weidner} {et~al.}(2004){Weidner}, {Kroupa}, \& {Larsen}}]{weidner04}
{Weidner}, C., {Kroupa}, P., \& {Larsen}, S.~S. 2004, \mnras, 350, 1503

\bibitem[{{Weidner} {et~al.}(2013){Weidner}, {Kroupa}, \&
  {Pflamm-Altenburg}}]{weidner_k_pa13}
{Weidner}, C., {Kroupa}, P., \& {Pflamm-Altenburg}, J. 2013, \mnras, 434, 84

\bibitem[{{Whitmore} {et~al.}(2007){Whitmore}, {Chandar}, \&
  {Fall}}]{whitmore07}
{Whitmore}, B.~C., {Chandar}, R., \& {Fall}, S.~M. 2007, \aj, 133, 1067

\bibitem[{{Whitmore} {et~al.}(2010){Whitmore}, {Chandar}, {Schweizer},
  {Rothberg}, {Leitherer}, {Rieke}, {Rieke}, {Blair}, {Mengel}, \&
  {Alonso-Herrero}}]{whitmore10}
{Whitmore}, B.~C., {Chandar}, R., {Schweizer}, F., {et~al.} 2010, \aj, 140, 75

\bibitem[{{Yamaguchi} {et~al.}(2001){Yamaguchi}, {Mizuno}, {Mizuno}, {Rubio},
  {Abe}, {Saito}, {Moriguchi}, {Matsunaga}, {Onishi}, {Yonekura}, \&
  {Fukui}}]{yamaguchi01}
{Yamaguchi}, R., {Mizuno}, N., {Mizuno}, A., {et~al.} 2001, \pasj, 53, 985

\bibitem[{{Zhang} \& {Fall}(1999)}]{zhang_fall99}
{Zhang}, Q. \& {Fall}, S.~M. 1999, \apj, 527, L81

\end{thebibliography}
\bibliographystyle{aa} 

\vspace{0.2 cm}
\noindent {\scriptsize This paper has been typeset from a \TeX/\LaTeX~file prepared by the author.}

\end{document}